# Structure, superconductivity, and magnetism in $Rb_{1-x}Fe_{1.6}Se_{2-z}S_z$

D. Croitori,[1] I. Filippova,[1] V. Kravtsov,[1] A. Günther,[2] S. Widmann,[2] D. Reuter,[2]
H.-A. Krug von Nidda,[2] J. Deisenhofer,[2] A. Loidl,[2] and V. Tsurkan[1,2]

[1]*Institute of Applied Physics, MD-2028 Chisinau, Republic of Moldova*
[2]*Experimental Physics V, Center for Electronic Correlations and Magnetism, University of Augsburg, 86135 Augsburg, Germany*

**Date: 04 September 2019**

**Abstract**. The single-crystal growth, stoichiometry, and structure of $Rb_{1-x}Fe_{2-y}Se_{2-z}S_z$ crystals with substitution of Se by S are reported. The variation of the magnetic and thermodynamic parameters of samples was studied by differential-scanning calorimetry, magnetic susceptibility, conductivity, and specific heat. The experimental results are discussed within a $T$-$z$ phase diagram, which includes vacancy-ordered and vacancy-disordered antiferromagnetic (AFM), superconducting (SC), and non-superconducting phases. The structural study revealed change in the local environment of the Fe tetrahedrons depending on substitution: a reduction of the Fe-Fe and Fe-Ch(chalcogen) bond lengths and a tendency for the six out of eight bond angles to approach values for a regular tetrahedron suggesting a reduction of structural distortions with substitution. With increasing substitution, a lowering of the superconducting transition temperature $T_c$ was observed; the percolation threshold for the SC state is located at the substitution $z = 1.2$. The SC state was found to coexist with the AFM state that persists in all samples independent of substitution. The temperature of the transition into the AFM state $T_N$ shows a monotonous decrease indicating a weakening of the AFM interactions with increasing substitution. The AFM phase exhibits an iron-vacancy-ordered structure below the structural transition at $T_s$. The temperature $T_s$ shows a non-monotonous variation: a decrease with increasing $z$ up to 1.3, followed by an increase for further increasing $z$. The suppression of the superconductivity with substitution is accompanied by a significant reduction of the density of states at the Fermi energy and a weakening of the electronic correlations in the studied system.

## I.    INTRODUCTION

In recent years, alkali-metal intercalated iron selenides $A_{1-x}Fe_{2-y}Se_2$ (with A = K, Rb, Cs) have attracted significant attention due to their unusual structural and electronic properties which are highly contrasting to other groups of Fe-based superconductors [1-3]. These materials show fascinating coexistence of insulating dominant phase $A_{0.8}Fe_{1.6}Se_2$ with iron-vacancy-ordered superstructure (known as 245 phase) and stripes of minority metallic iron-vacancy-free $A_{1-x}Fe_2Se_2$ phase (122 phase), which is believed to become superconducting at low temperatures. The insulating phase exhibits an antiferromagnetic (AFM) spin ordering with a Néel temperature $T_N$ above 500 K and a large local magnetic moment of 3.3 $\mu_B$ per Fe ion arranged in clusters of four iron ions [4]. Despite numerous studies of intercalated Fe chalcogenides using various local and macroscopic techniques, the interrelation between the AFM and SC phases is far from being well understood (see Refs. 3,5 and references therein). The critical temperature of the transition into the superconducting (SC) state $T_c$ of the order of 28-33 K for all members of the family $A_{1-x}Fe_{2-y}Se_2$ [6-14] is quite close to $T_c$ of 37 K for bulk FeSe under high pressure [15,16]. This significantly enhanced value of $T_c$ by external pressure compared to $T_c$ of 8 K for bulk FeSe at ambient pressure [17] suggests a connection of the SC parameters and local structural environment of Fe atoms thus indicating the importance of structural and electronic correlations for Fe chalcogenide superconductors. At the same time, the value of $T_c$ for $A_{1-x}Fe_{2-y}Se_2$ is considerably reduced compared to $T_c$ of 65-100 K reported recently for monolayers of FeSe [18,19]. With respect to pressure effects, the chemical pressure induced by doping or substitution is well known as an effective way to tune $T_c$ of Fe-based superconductors. For instance, the substitution of Se by Te in FeSe allows to enhance the value of $T_c$ for $FeSe_{0.5}Te_{0.5}$ up to 14 K [20,21]. However, the Fe-Se-Te system does not allow continuous variation of $T_c$ by substitution. Similarly, the critical temperature $T_c$ for all three $A_{1-x}Fe_{2-y}Se_2$ systems is only weakly dependent on variations of their composition within the existence range of superconductivity [5,14]. Moreover, the SC behavior in these materials is observed only in a narrow range of iron non-stoichiometry making it difficult to unravel the correlations between structural and electronic properties. An attempt to overcome these problems was demonstrated by Lei *et al.* [22] who utilized anion substitution in $K_xFe_{2-y}Se_2$. By replacement of S for Se, they were able to continuously tune the SC temperature up to a full suppression of superconductivity. The suppression of superconductivity was attributed to an increasing distortion of the Fe2-Se tetrahedron and increasing occupancy of the Fe1 site occurring simultaneously with growing substitution. Here Fe1 and Fe2 refer to the two different sites of the iron ions in the crystal structure described within the tetragonal $I4/m$ cell, which were assumed to be nearly empty and fully occupied, respectively. It was suggested that the increasing distortion leads to carrier localization and/or a decrease of the density of states at the Fermi energy [22]. Recently, an angle-resolved photoemission spectroscopy study has been performed on $Rb_{1-x}Fe_{2-y}Se_{2-z}S_z$ for three concentrations $z = 0$, 1, and 2 [23]. When moving from the non-superconducting phase ($z = 2$) to the SC phase ($z = 0$), a reduction of the bandwidth by a factor of two was found. The bandwidth was identified as a primary tuning parameter for superconductivity. Later on, terahertz time-domain spectroscopy studies of SC and metallic $Rb_{0.75}Fe_{1.6}Se_{2-z}S_z$ samples with more detailed substitution concentrations revealed a metal-to-insulator transition assisted with an orbital-selective Mott phase [24,25]. It was shown that the orbital-selective Mott transition shifts to higher temperatures with increasing S substitution indicating a reduction of correlations in the $d_{xy}$ channel that can account for the observed suppression of $T_c$ [24].

In the present paper, a complete macroscopic characterization of single crystalline samples of the $Rb_{1-x}Fe_{2-y}Se_{2-z}S_z$ studied earlier in [24] and additional samples with narrow substitutional steps ($z = 0$; 0.1; 0.25; 0.5; 1.0; 1.1; 1.2; 1.3; 1.4; 1.7, and 2.0) is reported. The results on the stoichiometry, structure, differential scanning calorimetry, magnetic susceptibility, conductivity, and specific heat are presented. We investigated the variation of structural and electronic properties of crystals with different substitution in order to establish the critical concentration for suppression of the SC state and to search for correlation effects using the advantage of the Rb-based system to form superconducting compositions with much smaller deviations of the iron stoichiometry from the 245 phase [14] compared to the related K-based system [22].

## II. EXPERIMENTAL DETAILS

Single crystals of the anion-substituted $Rb_{1-x}Fe_{2-y}Se_{2-z}S_z$ have been grown by the Bridgman method. The chemical composition of the samples was determined by electron-probe microanalysis (EPMA) applying wavelength dispersive x-ray spectroscopy (WDS) with a Cameca SX50 microprobe.

The single crystal x-ray diffraction was performed at room temperature with an Xcalibur E diffractometer equipped with a CCD area detector and a graphite monochromator utilizing MoK$_\alpha$ radiation. Samples for x-ray experiments were cut from large crystal pieces and protected with Paratone-N oil. Final unit cell dimensions were obtained and refined for the entire data set. After collection and integration, the data were corrected for Lorentz and polarization effects and for absorption by multi-scan empirical correction methods. The structures were refined by the full matrix least-squares method based on $F^2$ with anisotropic displacement parameters. All calculations were carried out by the programs SHELXL2014 [26,27]. Mixed Se/S sites were refined in a similar way. In each position, the Se and S atoms were constrained to have identical coordinates and thermal parameters.

Differential scanning calorimetry (DSC) measurements were performed using a PerkinElmer DSC-8500 system. The data were collected during temperature sweeps for heating and cooling with a rate of 5 K/min. The samples were encapsulated in standard Al crucibles. During the experiments, Ar or He gases were used as protecting media. The heat flow was normalized to the mass of the samples.

Magnetic characterization was performed using a commercial SQUID magnetometer (MPMS-5, Quantum Design) for temperatures between 1.8 K and 700 K, in external magnetic fields up to 5 T.

The resistivity and specific heat were measured with a Physical Property Measurement System (PPMS, Quantum Design) in a temperature range from 1.8 to 300 K.

### III. EXPERIMENTAL RESULTS AND DISCUSSION
#### A. Preparation conditions and composition analysis of samples

The preparation conditions and regimes of the single crystals growth by the vertical Bridgman method were similar to those for the non-substituted $Rb_{1-x}Fe_{2-y}Se_{2-z}S_z$ [14]. As starting materials, polycrystalline binary compounds FeSe and FeS, preliminary synthesized from the high-purity elements: Fe (99.99%), Se(99.999%), and S (99.999%), and elemental Rb (99.75%) were used. Handling of the reaction mixtures was performed in an argon box with a residual oxygen and water content less than 1 ppm. The starting materials were placed in double quartz ampoules, pumped to $10^{-3}$ mbar, and then closed. The ampoules were heated to a soaking temperature of 1070 $^o$C. The soaking time was 5 h. Then, the ampoules were pulled down in the temperature gradient of 300 $^{o}$C with a rate of 3 mm/h. The composition of the starting mixtures for batches with different substitution levels is given in Table 1.

The concentration of the elements in the studied samples was measured on freshly cleaved samples. The EPMA data are also given in Table 1. All values are averaged over multiple (ten to twenty) measured spots with an area of $80 \times 60$ µm$^2$. The errors in determination of the absolute concentrations of the elements were less than 1.5% for Fe, 2% for Se and S, and 5% for Rb. The concentrations of Rb and Fe were calculated normalizing the sum of Se+S to two atoms per-formula unit. The EPMA analysis did not reveal any essential deviations in the S to Se ratio from the starting stoichiometry for all grown batches. The concentration of Fe in the samples from different batches was close to 1.6 indicating compositions with a Fe-vacancy corresponding to the 245 stoichiometry. We note that the deviations from the 245 stoichiometry in the Rb$_{1-x}$Fe$_{2-y}$Se$_{2-z}$S$_z$ system are much smaller than in K$_x$Fe$_{2-y}$Se$_{2-z}$S$_z$ where significant variations of the Fe content from 1.44 to 1.72 on increasing substitution from $z = 0$ to $z = 2$ were reported [22].

An important observation concerns the microstructure of the studied samples. As reported earlier for non-substituted samples ($z = 0$) in Ref. 28, the presence of two different phases is easily distinguished in an optical microscope with µm-size metallic stripes embedded into the AFM 245 phase (see also Fig. 1SM of Supplemental Material). Under high resolution conditions, it was possible to determine the composition of these stripes, which corresponds to Rb$_{0.705(25)}$Fe$_{2.017(10)}$Se$_2$. Within experimental uncertainty, this corresponds to a Fe vacancy-free and a Rb-deficient 122 phase. This result correlates well with those obtained by other techniques. However, a much higher Rb/Fe ratio (0.7/2) for stripes was revealed compared to that in the previous neutron diffraction study (0.6/2.2) [29] and nuclear magnetic resonance (0.3/2) [30]. Since the WDS analysis is an absolute and accurate method of compositional determination, we assume that our result can be considered to be most reliable. It was further observed that for samples even with the lowest substitution $z = 0.1$, it was not possible to detect any stripe structure in the µm range (see Fig. 1SM of the Supplemental Material). At the same time, the presence of the AFM and metallic non-magnetic phases was detected in all samples by Mössbauer experiments [31] indicating that the phase separation in the anion substituted crystals is realized on a significantly lower length scale.

Table 1. Starting mixtures and real compositions of selected samples of
$Rb_{1-x}Fe_{2-y}Se_{2-z}S_z$ as determined by EPMA analysis

| Batch and sample label | Substitution $z$ | Starting mixture | Concentration of the elements | | | |
|---|---|---|---|---|---|---|
| | | | Rb $(1-x)$ | Fe $(2-y)$ | Se $(2-z)$ | S $(z)$ |
| BR16 | 0 | 0.8Rb+2FeSe | 0.748(27) | 1.593(16) | 2.000(19) | - |
| BR16_05 | 0 | 0.8Rb+2FeSe | 0.736(40) | 1.611(14) | 2.000(30) | - |
| BR28 | 0 | 0.8Rb+2FeSe | 0.786(39) | 1.612(22) | 2.000(28) | - |
| BR100 | 0.1 | 0.8Rb+1.9FeSe+0.1FeS | 0.750(33) | 1.596(13) | 1.905(22) | 0.095(2) |
| BR99 | 0.25 | 0.8Rb+1.75FeSe+0.25FeS | 0.739(26) | 1.592(16) | 1.752(22) | 0.248(7) |
| BR96_le | 0.5 | 0.8Rb+1.5FeSe+0.5FeS | 0.734(25) | 1.597(27) | 1.511(20) | 0.489(13) |
| BR96_1 | 0.5 | 0.8Rb+1.5FeSe+0.5FeS | 0.734(24) | 1.603(26) | 1.507(20) | 0.493(13) |
| BR80 | 1.0 | 0.8Rb+FeSe+FeS | 0.765(23) | 1.605(19) | 1.017(20) | 0.983(18) |
| BR87 | 1.0 | 0.8Rb+FeSe+FeS | 0.764(27) | 1.595(16) | 0.998(21) | 1.002(16) |
| BR82 | 1.1 | 0.8Rb+0.9FeSe+1.1FeS | 0.844(32) | 1.585(20) | 0.922(23) | 1.079(24) |
| BR101_1 | 1.4 | 0.8Rb+0.6FeSe+1.4FeS | 0.802(15) | 1.620(14) | 0.634(25) | 1.366(16) |
| BR101_Ro1 | 1.4 | 0.8Rb+0.6FeSe+1.4FeS | 0.791(36) | 1.610(17) | 0.650(23) | 1.350(8) |
| BR102_1 | 1.7 | 0.8Rb+0.3FeSe+1.7FeS | 0.822(21) | 1.585(18) | 0.312(16) | 1.688(15) |
| BR97_optics | 2.0 | 0.8Rb+2FeS | 0.787(16) | 1.595(11) | - | 2.000(12) |
| BR97_1 | 2.0 | 0.8Rb+2FeS | 0.735(16) | 1.611(17) | - | 2.000(24) |

## B. Structural study

The x-ray single-crystal structure analysis of the experimental pattern of $Rb_{1-x}Fe_{2-y}Se_{2-z}S_z$ reveals the presence of a tetragonal cell with large lattice parameters $a(b)$~19 Å, $c$~14 Å for all substitutions. The initial refinement of the crystal structure was performed within the space group $I4/m$ with a large $5 \times 5 \times 1$ supercell. Structural data and details of the structural refinement for three selected compositions (with $z = 0$, 1, and 2) are given in Tables 2(a) and (b). In the space group $I4/m$, there are seven different Fe sites: four are fully occupied (Fe3, Fe4, Fe5, Fe6), two (Fe2, Fe7) are partially occupied, and one (Fe1) is vacant (see Fig. 2SM of the Supplemental Material). Within this model, it is indeed possible to get a fully vacant position for Fe1 in contrast to the "formally fully vacant" Fe1 in the $\sqrt{5} \times \sqrt{5} \times 1$ cell. Structural data and details of the structural refinement for all substitutions performed within the $5 \times 5 \times 1$ supercell are given in Table 1SM of the Supplemental Materials.

It is worth mentioning that the structural refinement within a large $5 \times 5 \times 1$ supercell was reported earlier by Zavalij *et al.* [32] for $K_{1-x}Fe_{2-y}Se_2$ and $Cs_{1-x}Fe_{2-y}Se_2$ crystals. Their initial structural analysis was done in the space group $I4/mmm$. In that group, there are six different Fe sites: four are fully occupied and one is fully vacant, while the last one is partially occupied for about 40%. However, those authors noticed the incompatibility of the space group $I4/mmm$ with the experimental data on powder neutron diffraction [3], and the structural data were further interpreted within the supercell $\sqrt{5} \times \sqrt{5} \times 1$ with $a(b) \sim 8.7$ Å with only two different Fe sites.

It must be noted that the structural solution within the cell $\sqrt{5} \times \sqrt{5} \times 1$ describes an averaged structure of our $Rb_{1-x}Fe_{2-y}Se_{2-z}S_z$ samples. It neglects about 30% of the observed experimental intensity with a fully regular diffraction pattern. This can be concluded by comparing the reciprocal lattice plots for $5 \times 5 \times 1$ and $\sqrt{5} \times \sqrt{5} \times 1$ cells for one of the samples with $z = 1$ shown in Figs. 3SM (a) and 3SM (b) of the Supplemental Materials.

Table 2 (a). Structural data and details of structural refinement for samples with substitution $z = 0$, 1, and 2 within space group $I4/m$ for $5 \times 5 \times 1$ supercell.

| $z$ | 0 | 1 | 2 |
|---|---|---|---|
| Formula weight | 314.66 | 254.38 | 218.59 |
| $a = b$ (Å) | 19.677(1) | 19.286(1) | 18.935(1) |
| $c$ (Å) | 14.585(2) | 14.352(1) | 14.039(1) |
| Volume (Å$^3$) | 5646.7(7) | 5338.3(4) | 5033.2(6) |
| Reflections collected /unique | 45872/2730 $R_{int} = 0.2419$ | 52362/3352 $R_{int} = 0.1757$ | 36487/2437 $R_{int} = 0.1457$ |
| $GooF$ | 1.007 | 1.002 | 1.022 |
| $R_1$, $wR_2$ [$I > 2\sigma(I)$] | 0.0594, 0.1418 | 0.0863 0.1673 | 0.0794 0.1634 |

Table 2 (b). Atomic coordinates ($x$, $y$, $z$) and site-occupation factors (sof) for Fe ions within space group $I4/m$ for $5 \times 5 \times 1$ supercell for samples with substitution $z = 0$, 1, and 2.

| | Fe1 | Fe2 | Fe3 | Fe4 | Fe5 | Fe6 | Fe7 | Substitution |
|---|---|---|---|---|---|---|---|---|
| $x$ | 0.5 | 0.206(2) | 0.5023(2) | 0.3031(2) | 0.4979(3) | 0.4023(3) | 0.3965(2) | $z = 0$ |
| $y$ | 0 | 0.111(1) | 0.4026(2) | 0.1969(2) | 0.2039(2) | 0.1017(2) | 0.2974(2) | |
| $z$ | 0.75 | 0.748(1) | 0.7538(3) | 0.7488(5) | 0.7473(2) | 0.7480(6) | 0.7479(4) | |
| sof | 0 | 0.104(2) | 1.0 | 1.0 | 1.0 | 1.0 | 0.896(2) | |
| $x$ | 0.5 | 0.2021(2) | 0.4999(1) | 0.3033(1) | 0.4999(1) | 0.4021(1) | 0.3945(2) | $z = 1$ |
| $y$ | 0 | 0.1051(2) | 0.4016(1) | 0.1969(1) | 0.2046(1) | 0.0981(1) | 0.2987(2) | |
| $z$ | 0.75 | 0.7488(3) | 0.75391(1) | 0.7504(1) | 0.7465(1) | 0.7504(2) | 0.7488(3) | |
| sof | 0 | 0.504(2) | 1.0 | 1.0 | 1.0 | 1.0 | 0.496(2) | |
| $x$ | 0.5 | 0.2007(2) | 0.4996(1) | 0.3034(1) | 0.4997(1) | 0.4024(1) | 0.3938(2) | $z = 2$ |
| $y$ | 0 | 0.1058(2) | 0.4000(1) | 0.1970(1) | 0.2056(1) | 0.972(1) | 0.3002(2) | |
| $z$ | 0.75 | 0.7513(2) | 0.7539(1) | 0.7503(1) | 0.7460(1) | 0.7499(2) | 0.7513(2) | |

| sof | 0 | 0.542(2) | 1.0 | 1.0 | 1.0 | 1.0 | 0.458(2) | |

In Fig. 1, the variations of the lattice parameters $a$ and $c$ with substitution are presented. Both parameters show a linear decrease with increasing sulfur content following Vegard`s law and indicating the formation of continuous solid solutions in this system due to substitution of S for Se. This fact together with the absence of any additional changes in the lattice symmetry with substitution indicates statistical substitution of Se by the S ions in the anion positions.

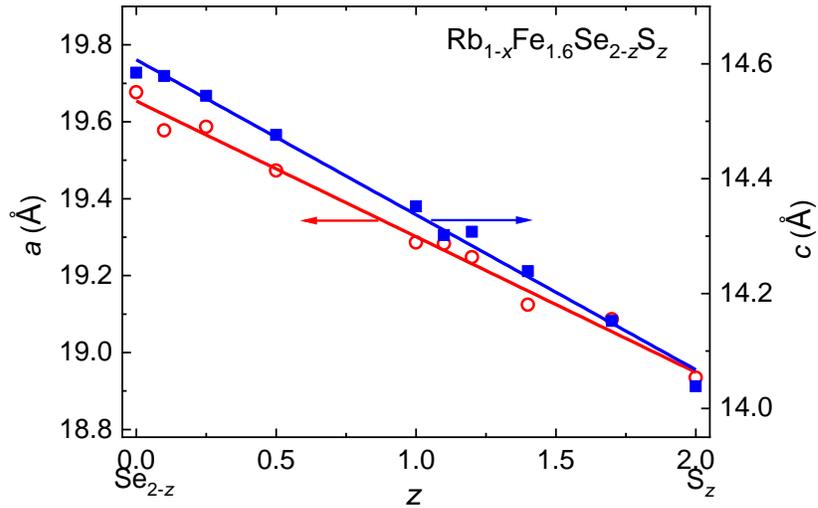

FIG. 1. Variation of the lattice parameters $a$ and $c$ with substitution $z$ in $Rb_{1-x}Fe_{2-y}Se_{2-z}S_z$.

In addition to the initial structural model, the structural refinement of all samples was also performed within the generally accepted $\sqrt{5} \times \sqrt{5} \times 1$ cell, in the space group $I4/m$. This allows a direct comparison of the structural data for our $Rb_{1-x}Fe_{2-y}Se_{2-z}S_z$ system with the related data for $K_xFe_{2-y}Se_{2-z}S_z$ in [22]. The respective data are given in Table 2SM of the Supplemental Material. In the $\sqrt{5} \times \sqrt{5} \times 1$ cell, all constituent elements have two different crystallographic positions (see Fig. 2) with different occupancies for Fe and Rb sites. The site occupancy for the Fe2 ion in a general position ($x, y, z$) is close to 0.93, while that of the Fe1 ion in a special position (0.5, 0, 0.25) is close to 30 %. The occupancy of both Fe sites in $Rb_{1-x}Fe_{2-y}Se_{2-z}S_z$ exhibits only an insignificant change with substitution (see upper panel of Fig. 6 below). The Rb sites are also partially occupied. The occupancy of the Rb sites (see Table 3SM of Supplemental Material) shows a non-monotonous change with substitution, which probably has to be attributed to variations of the concentration of Rb in the samples, being particularly high for samples with substitution $z = 1.1$ (see Table 1). The reason of this variation of Rb concentration is unclear at the moment.

The local tetrahedral environment of Fe2 ion consists of three nearly equivalent Se2(S2) neighbors and of one Se1(S1) ion, while the local environment of Fe1 ion consists of four equivalent Se2(S2) ions as shown in Fig. 2(b). Four Fe2 ions form clusters with shorter intra-cluster and larger inter-cluster distances Fe2-Fe2.

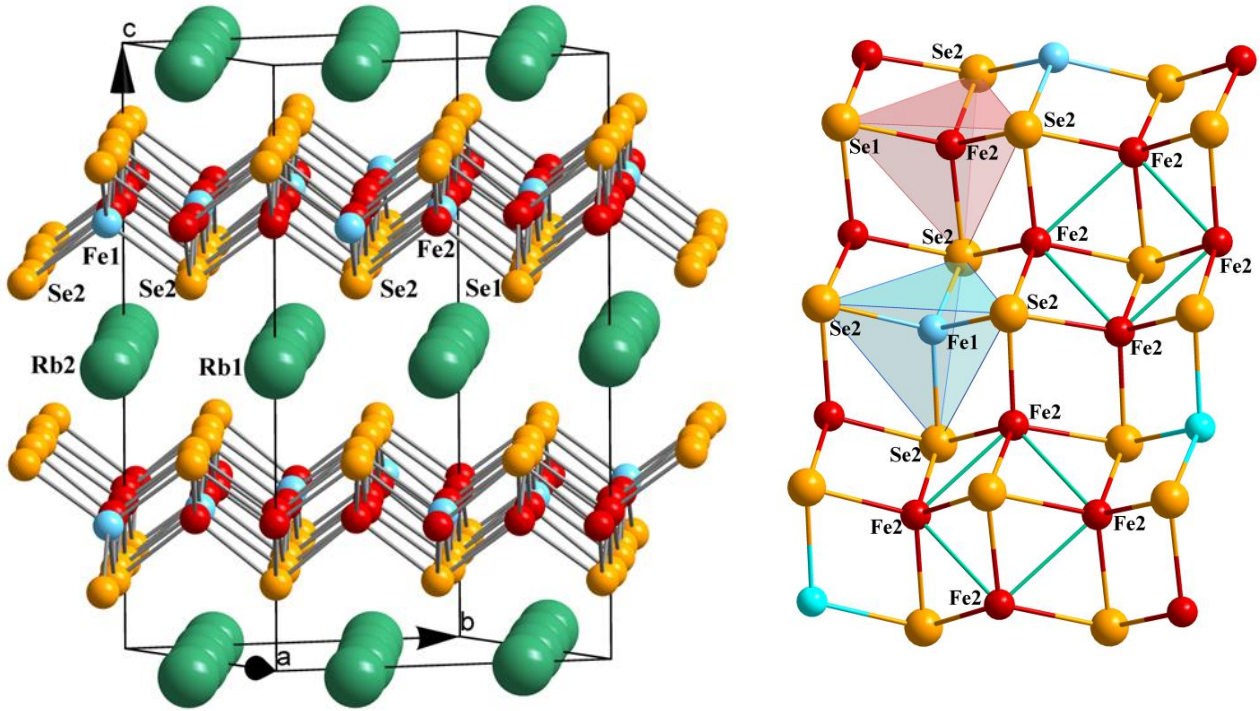

FIG. 2 (a). Crystal structure of $Rb_{1-x}Fe_{2-y}Se_{2-z}S_z$ for $z$=0. Fe1 ions are in position (0.5, 0, 0.25), Fe2 in position (*x, y, z*); Rb1 (0, 0, 0.5), Rb2 (*x, y*, 0.5), Se1(S1) in position (0.5, 0.5, *z*), and Se2(S2) in (*x, y, z*).

FIG. 2 (b). Schematic view of different Fe tetrahedrons: Fe2 with three neighbors Se2(S2) and one Se1(S1); Fe1 with four equivalent neighbors Se2(S2). Clusters of four Fe2 ions are marked by rectangles.

To get insight into the lattice distortions and their evolution with substitution, we analyzed the bond distances between the Fe ions, between the Fe ions and chalcogens (Ch = Se or S), as well as the bond angles in Fe2 and Fe1 tetrahedrons. A monotonous decrease of the bond distances Fe-Ch for both tetrahedrons was found (see Fig. 3 (a) and Table 4SM of the Supplemental Material), which is in agreement with that anticipated for the observed decrease of the unit cell parameters *a* and *c* with substitution (Fig. 1). However, it was noticed that the rate of the decrease of all Fe-Ch distances is notably higher for substitution $z$ above 1.2. For $z \leq 1.2$, the bond distances in Fe2-Ch1 tetrahedron are larger than in Fe1-Ch2 tetrahedron, but, for $z > 1.2$, they become lower. At the same time, for $z \leq 1.2$, the ratio of three nearly equivalent bond distances Fe2-Ch2 to distance Fe1-Ch2 changes slightly, whereas above $z = 1.2$, it starts to decrease essentially (Fig. 3 (b)). The Fe1-Fe2 distance and the inter-cluster Fe2-Fe2 distance show a much stronger decrease with the substitution than the intra-cluster Fe2-Fe2 distance (Fig. 4 (a)). The observed variations of the Fe-Fe distances with substitution are in good agreement with those reported for $K_x Fe_{2-y} Se_{2-z} S_z$ [22]. It must be noted that the ratios of the Fe1-Fe2 distance to the inter-cluster distance Fe2-Fe2 and of the intra-cluster distance Fe2-Fe2 to the inter-cluster distance Fe2-Fe2 exhibit an opposite trend with substitution (Fig. 4 (b)) intercepting in the range close to $z = 1.2$. As will be shown below, the superconductivity in the samples vanishes just above $z = 1.2$. It is also worth mentioning that, with increasing substitution, the ratio of the intra-cluster to inter-cluster Fe2-Fe2 distances increases and approaches unity suggesting a more regular in-plane structural arrangement.

In Figs. 5 (a-c), the variations of the bond angles for Fe1 and Fe2 tetrahedrons with substitution are presented. In the Fe1 tetrahedron, four large angles $\alpha_1$ and two small angles $\alpha_2$ exhibit a tendency to approach the ideal angle of 109º47' on increasing substitution up to $z = 2.0$ (see Fig. 5 (a)). A similar trend is found for the other four angles $\alpha_3$, $\alpha_4$, $\alpha_{5,}$ and $\alpha_6$ in the Fe2 tetrahedron, while the remaining two angles $\alpha_7$ and $\alpha_8$ of this tetrahedron increase and, respectively, decrease with substitution (see Fig. 5 (c)). An analysis of the regularity of Fe tetrahedrons by comparing the sum of the angles at the Ch1 and Ch2 vertices has shown that with increasing substitution from $z = 0$ to $z = 2$, the Fe2 tetrahedron becomes more regular indicating decreasing distortions (see Table 5SM of the Supplemental Material). This fact is in a clear contrast to that reported for the $K_x Fe_{2-y} Se_{2-z} S_z$ system, where increasing distortions with S substitution were observed and were suggested to destroy the SC state [22].

Looking for a possible optimization of the structural parameters, we have analyzed the variations of the anion height with substitution. Both anion heights of the Ch1 to the Fe2 plane and of the Ch2 to the Fe1-Fe2 plane were found to exhibit a continuous decrease with substitution as shown on the lower panel of Fig. 6. This behavior is also in a strong contrast to that observed in $K_x Fe_{2-y} Se_{2-z} S_z$ [22].

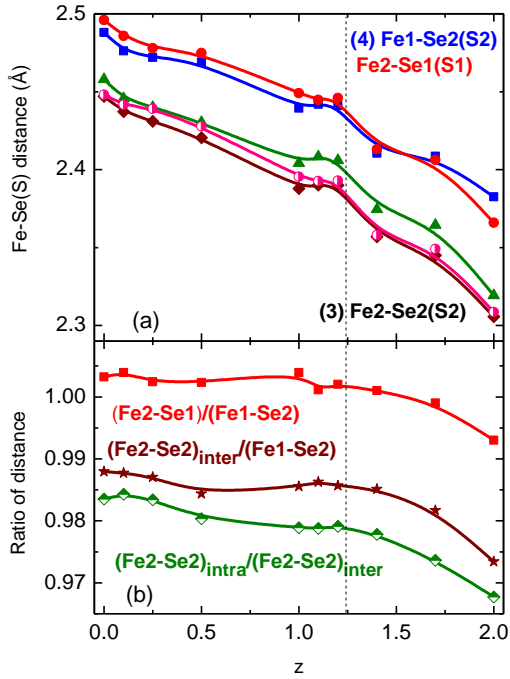

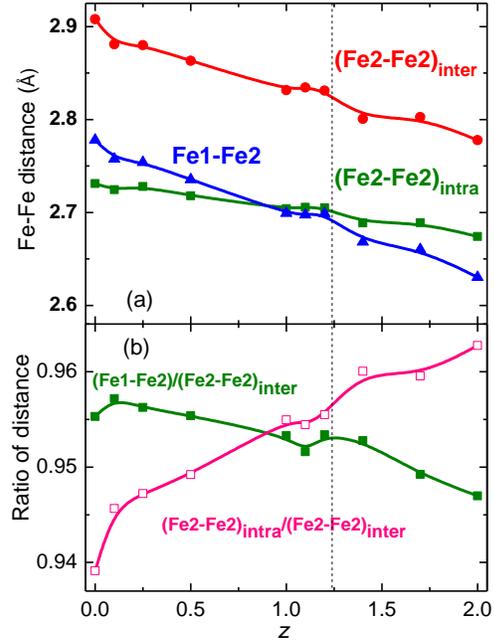

FIG. 3. Variations with substitution. Frame (a): four equivalent bond distances Fe1-Ch2, four equivalent distances Fe2-Ch1, and three bond distances Fe2-Ch2. Frame (b) ratio of bond distance Fe2-Ch1 to Fe1-Ch2, intercluster distances Fe2-Ch2 to Fe1-Ch2, intracluster distances Fe2-Ch2 to intercluster Fe2-Ch2. Vertical dashed line separates superconducting samples from non-superconducting ones.

FIG. 4. Variations with substitution. Frame (a): inter-cluster Fe2-Fe2 distances, intra-cluster Fe2-Fe2 distances, and Fe1-Fe2 distances. Frame (b) ratio of Fe1-Fe2 to inter-cluster Fe2-Fe2 distance, intra-cluster Fe2-Fe2 to inter-cluster Fe2-Fe2 distance.

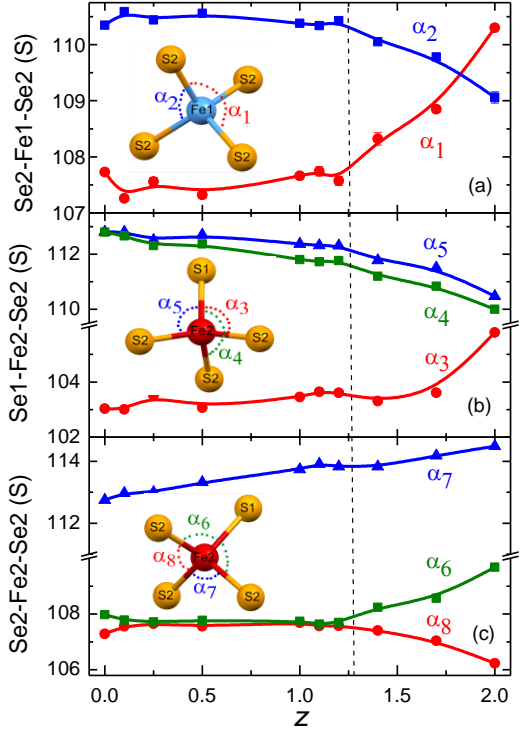

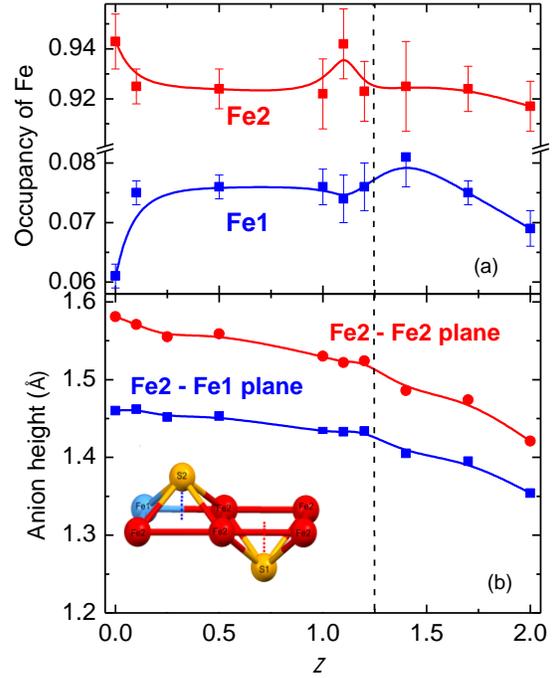

FIG. 5. Variation with substitution of bond angles (in degrees). Frame (a): Fe1 tetrahedron (two angles $\alpha_1$ and four angles $\alpha_2$. Frames (b) and (c): Fe2 tetrahedron (angles from $\alpha_3$ to $\alpha_8$). Vertical dashed line separates superconducting samples from the non-superconducting ones.

FIG. 6. Variations with substitution: of occupancy of Fe1 and Fe sites (frame a), and of anion-height from Ch (Se,S) to Fe2-Fe1 plane and from Ch to Fe2-Fe2 plane (frame b).

## C. Differential scanning calorimetry

Figs. 7 (a-d) show the DSC results for several selected $Rb_{1-x}Fe_{2-y}Se_{2-z}S_z$ crystals with different substitution. The full set of the DSC data for the studied crystals is given in the Supplemental Material (see Figs. 4SM (a-h)). The DSC signal corresponds to a difference in heat required to increase the sample temperature with respect to the reference (empty Al crucible). For the non-substituted sample (Fig. 7 (a), $z = 0$) on increasing temperatures from 300 K to 600 K, three clear anomalies were recorded. The temperature positions of these anomalies were very close to those found for the as grown $Rb_xFe_{2-y}Se_2$ single crystals studied by the DSC technique in [29, 32]. The neutron diffraction studies, also performed in [29], allowed assigning the anomaly in the DSC signal at the highest temperature $T_s$ to a structural transition of the dominant 245 phase from the vacancy-disordered state into the state with ordering of Fe vacancies. The anomaly at $T_N$ at intermediate temperature was assigned to a transition of the dominant 245 phase into the AFM state. The anomaly at the lowest temperature at $T_p$ was attributed to a phase-separation temperature, where the Rb-deficient 122 phase segregates from the 245 phase [29].

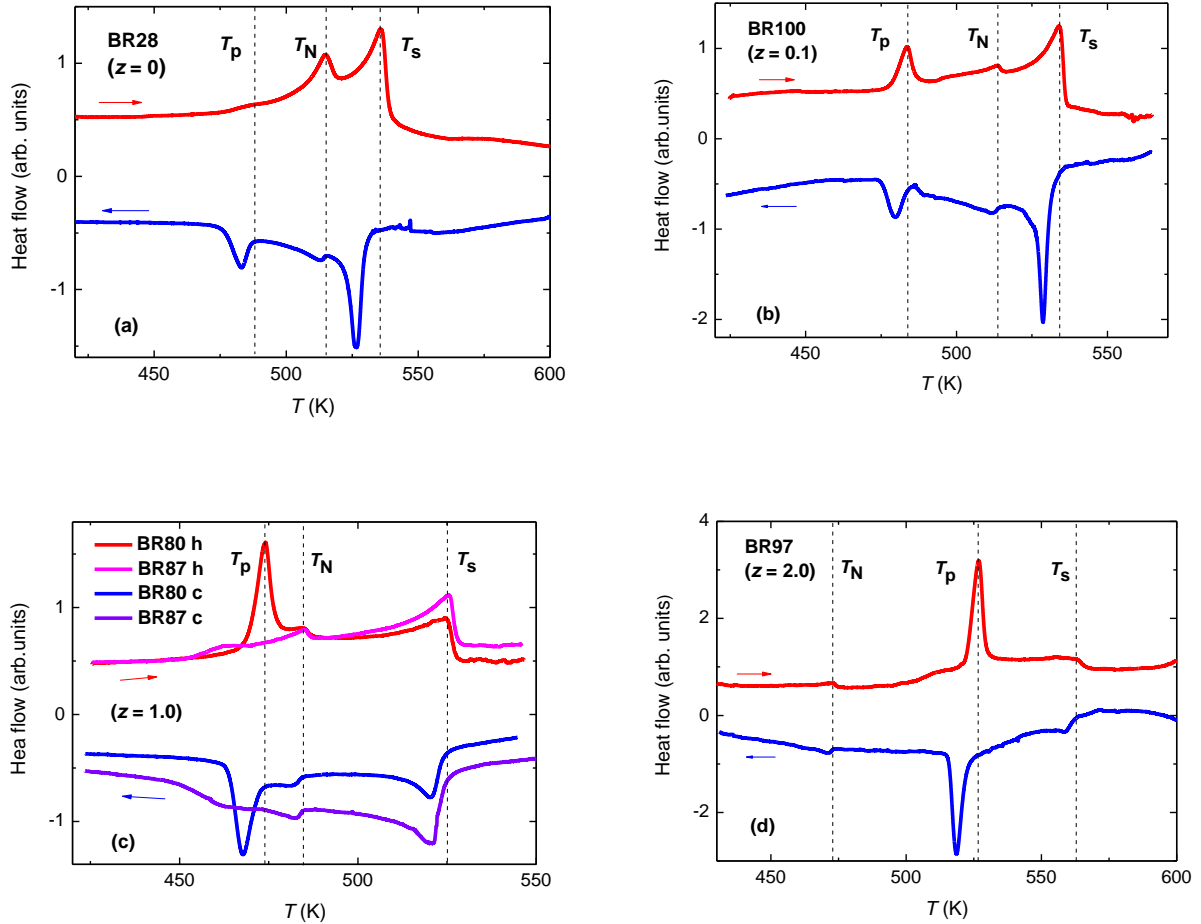

FIG. 7. Temperature dependence of DSC signals for $Rb_{1-x}Fe_{2-y}Se_{2-z}S_z$ crystals with different substitution (frame a: $z = 0$, frame b: $z = 0.1$, frame c: $z = 1$, frame d: $z = 2$). Red curves show data on heating, blue ones on cooling. Vertical dashed lines mark phase transformations on heating.

Inspection of the observed anomalies in the DSC signal for our $Rb_{1-x}Fe_{2-y}Se_{2-z}S_z$ crystals reveals that the anomalies at $T_s$ and $T_p$ exhibit significant hysteresis (up to 10 K) on cooling and heating cycles indicating that they are related to the first-order structural transformations. The anomaly at $T_N$ shows the smallest hysteresis (2 K) as expected for the second-order magnetic transformation.

The structural anomaly at $T_p$ was found to exhibit a complex appearance. In samples with substitution $z \leq 1.0$ this anomaly was discernible in the DSC signal as a small step or pronounced maximum on heating (see Figs. 7 (a) and (b)). In the samples with the step-like anomaly at $T_p$ it was better evidenced as pronounced minimum on cooling. Even for samples from different batches but with the same substitution $z = 1.0$, the anomaly at $T_p$ showed a completely different appearance on heating, although the other two anomalies at $T_N$ and $T_s$ were quite similar both for heating and cooling cycles (see Fig. 7 (c)). Such a distinct behavior of the anomaly at $T_p$ can be probably attributed to different distribution of the SC minority phase in the bulk which is affected by details of cooling process during the crystal growth and by the heat treatment during the DSC measurements. For the intermediate range of substitution $1.1 < z < 1.3$, the anomaly at $T_p$ was hardly detectable (see Fig. 4SM (e)). This can be understood as due to a reduction of amount of the minority SC phase that takes place within this range of substitution. For samples with substitution $z \geq 1.3$, the intensity of the anomaly at $T_p$ in the DSC signal becomes much higher than the intensity of the anomalies at $T_N$ and $T_s$ (see Fig. 7 (d) and Fig. 4SM (f-h)). The observed strong hysteresis on heating and cooling still allows associating this anomaly with the structural transformation. However, significantly increased enthalpy of this transition compared to those of the other two anomalies at $T_N$ and $T_s$ for samples with $z \geq 1.3$ suggests different origin of this phase compared to SC phase present in samples with lower substitution.

It was further observed that, while the anomaly at $T_s$ is well separated from the anomalies at $T_p$ and $T_N$ for the whole range of substitution, the values of $T_p$ and $T_N$ in the range $1.1 < z < 1.4$ become close to each other making it difficult to distinguish the anomaly at $T_N$ due to its lower intensity in the DSC curve than that of the anomaly at $T_p$. Therefore, to have an additional proof of the origin of the anomalies at $T_N$ and $T_p$, susceptibility studies of our samples in the temperature range 300-600 K were carried out. The respective data are given in the Supplemental Material (Figs. 5SM (a-e)). All three anomalies revealed in the DSC measurements were distinguishable and identified as clear steps or change of slope in the susceptibility or inverse susceptibility data. A good agreement of the phase-transition temperatures $T_s$, $T_p$, and $T_N$ was observed in DSC and magnetic-susceptibility. A comparison of the hysteretic behavior of these anomalies allowed getting reliable assignment of the structural and magnetic transformations.

Fig. 8 shows the variations of the temperatures of the structural and magnetic transformations with substitution as derived from the DSC data on heating. The structural transitions exhibit a non-monotonic change with substitution with a small decrease of $T_s$ and $T_p$ in a range of $0 < z < 1.2$ followed by a strong increase of their values for $z > 1.3$. At the same time, $T_N$ changes monotonously decreasing from 515 K (for $z = 0$) to 472 K (for $z = 2$).

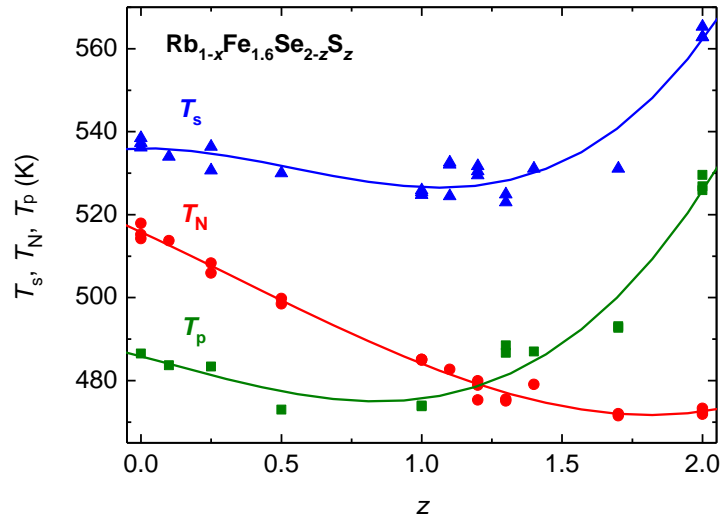

FIG. 8. Concentration dependence of the structural transformation temperatures $T_s$, and $T_p$, and of the magnetic transformation $T_N$ for $Rb_{1-x}Fe_{2-y}Se_{2-z}S_z$. Data are taken from DSC curves measured on heating. Solid lines are guides to the eye.

### D. Magnetic susceptibility and hysteresis

Figs. 9 (a) and 9 (b), respectively, present the temperature dependencies of the magnetic susceptibility, $\chi_\parallel$, for superconducting and non-superconducting samples with different substitutions measured in a magnetic field $H = 10$ kOe applied parallel to the $c$ axis. For the substituted samples, $\chi_\parallel$ shows a non-linear growth with temperature similar to pure $Rb_{0.8}Fe_{1.6}Se_2$ [14]. With growing the substitution $z$ up to 1.2, the susceptibility exhibits an insignificant increase in the value just above the SC transition. For the non-superconducting samples with $z \geq 1.3$, the susceptibility shows very similar over-all temperature dependence as for samples with $z \leq 1.2$ above the SC transition. In addition, in Fig. 9 (b), the susceptibility $\chi_\perp$ vs. temperature is shown for the sample with $z = 2$ measured in a configuration with the magnetic field applied perpendicular to the $c$ axis. Except for the temperatures below 50 K, the susceptibility $\chi_\perp$ is significantly enhanced and shows very little change with temperatures. A similar behavior of the temperature-dependent susceptibility $\chi_\perp$ with temperature was found for all samples studied in the course of these experiments. Such a behavior of $\chi_\perp$ and $\chi_\parallel$ is characteristic for an anisotropic antiferromagnet with the $c$ axis being the direction of alignment of the spins. Thus, the anisotropic antiferromagnetism observed in all studied samples is a distinct feature for the whole $Rb_{1-x}Fe_{2-y}Se_{2-z}S_z$ system.

Fig. 10 presents the temperature dependent zero-field cooled (ZFC), $\chi_{ZFC}$, and field-cooled (FC), $\chi_{FC}$, susceptibilities for superconducting samples with different substitutions measured in a field $H = 10$ Oe applied parallel to the $c$ axis. The value of the FC susceptibility (Meissner effect) is small due to strong pinning effect. At the same time, the value of the ZFC susceptibility indicates a 100% shielding effect for the samples with $z$ up to 1.2. With increasing substitution from 0 to 1.2, a continuous reduction of the superconducting transition temperature from 32.4 K to 10 K is observed, however, with a non-monotonous change at the substitution level of 1/8. This sample with $z = 0.25$ shows a lower transition temperature $T_c$ of 25 K than the sample with a higher substitution $z = 0.5$ with $T_c = 28$ K. The transition into the SC state for the samples with the substitution up to $z = 1.0$ is rather sharp. No broadening of the transition width for this substitution range was observed compared to non-substituted samples (again with exception of the sample with $z = 0.25$). The sample with $z = 1.2$ exhibits the lowest SC transition temperature close to 10 K.

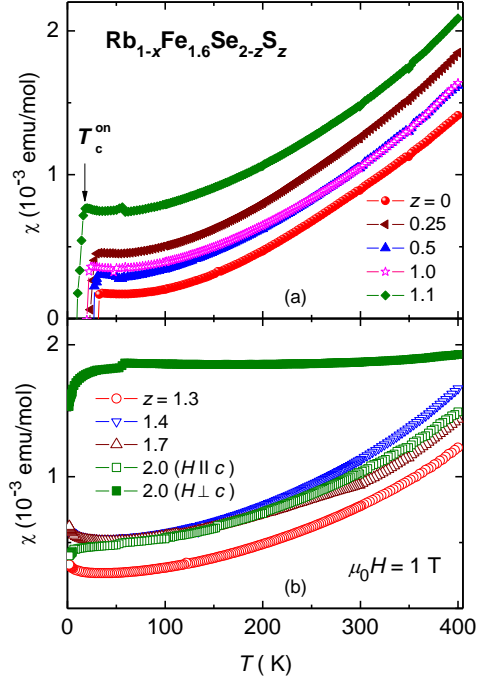

FIG. 9. Temperature dependent susceptibility $\chi_\parallel$ for superconducting (frame a) and non-superconducting (frame b) $Rb_{1-x}Fe_{2-y}Se_{2-z}S_z$ samples measured in an external magnetic field of 1 T applied along $c$ axis. Arrow indicates the SC transition temperature for the sample with $z = 1.1$. Additionally, the susceptibility $\chi_\perp$ measured in magnetic field applied perpendicular to the $c$ axis is shown for the sample with $z = 2$.

Fig. 11 presents magnetization hysteresis loops for SC samples measured at 2 K with the magnetic field $H$ applied along the $c$ axis. The diamagnetic response (i.e. the area of the loop) of the samples with the substitution range up to $z = 1.0$ (except for the sample with $z = 0.25$) is very similar to that of the non-substituted sample ($z = 0$). However, compared to the sample with $z = 0$, no fishtail effect is observable in the hysteresis loop for the substituted samples, even for a minor substitution $z = 0.1$. This indicates a significant change of the flux dynamics that occurs with substitution, which probably can be attributed to a difference in spatial distribution of the SC phase in the samples. As already mentioned, changes of the spatial distribution of the SC phase can be concluded from the change of microstructure (see Fig. 1SM of the Supplemental Material). A lower diamagnetic response of the sample with $z = 1.1$ may result from $z$ approaching the critical range of suppression of the SC state in this system, and concomitantly from the reduction of the amount of SC phase and of the modification of its distribution in the bulk. The reason of a strong reduction of the diamagnetic response for the sample with $z = 0.25$ is unclear and this anomalous behavior deserves further studies.

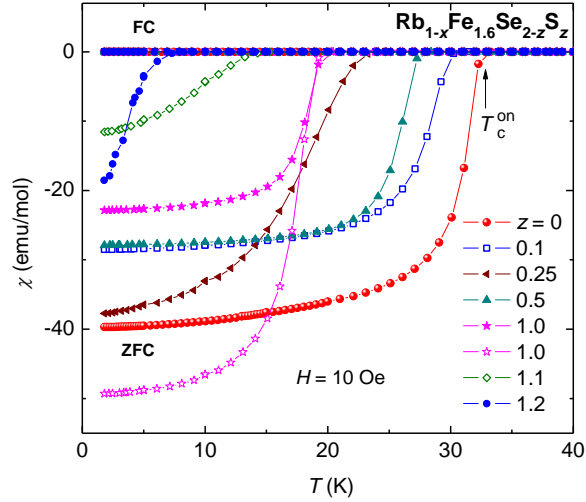

FIG. 10. Temperature dependent susceptibilities (ZFC and FC) for different superconducting $Rb_{1-x}Fe_{2-y}Se_{2-z}S_z$ samples measured in an external magnetic field of 10 Oe applied along $c$ axis. Arrow indicates the temperature of the onset of superconducting transition for the sample with $z = 0$.

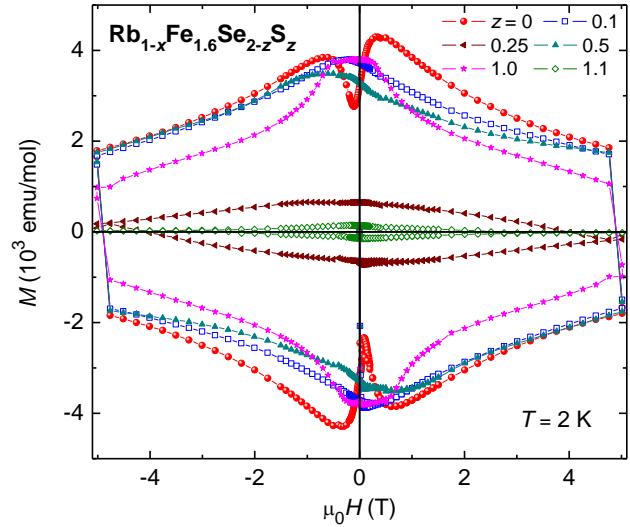

FIG. 11. Hysteresis loops measured at 2 K with the magnetic field applied along $c$ axis for different superconducting $Rb_{1-x}Fe_{2-y}Se_{2-z}S_z$ samples.

Thus, our magnetic studies revealed that the percolation threshold for the appearance of superconductivity in $Rb_{1-x}Fe_{2-y}Se_{2-z}S_z$ crystals is placed between $z = 1.2$ and $1.3$. It differs essentially from the $K_xFe_{2-y}Se_{2-z}S_z$ system where the SC state extends up to a substitution $z = 1.6$ [22].

### E. Resistivity

Figs. 12 and 13 show the temperature dependent resistivity of superconducting and non-superconducting samples, respectively. The resistivity for both types of samples exhibits non-monotonic temperature dependence with semiconductor-like behavior at high temperatures, a broad maximum at a characteristic temperature $T_m$ on decreasing temperature followed by a metallic-like behavior below $T_m$. The temperature $T_m$ shows a general trend to lower values with increasing substitution, however, with exception of $z = 0.25$ and $1.4$. In fact, such non-monotonic variation of $T_m$ with substitution is hard to understand; still we have registered a higher value of $T_m$ for samples with a lower residual resistivity in the normal state. Only the residual resistivity in the normal state for samples with substitution $z \leq 1.2$ shows a continuous increase with increasing sulfur content suggesting a decrease of the density of states at the Fermi level and/or an increase of disorder scattering.

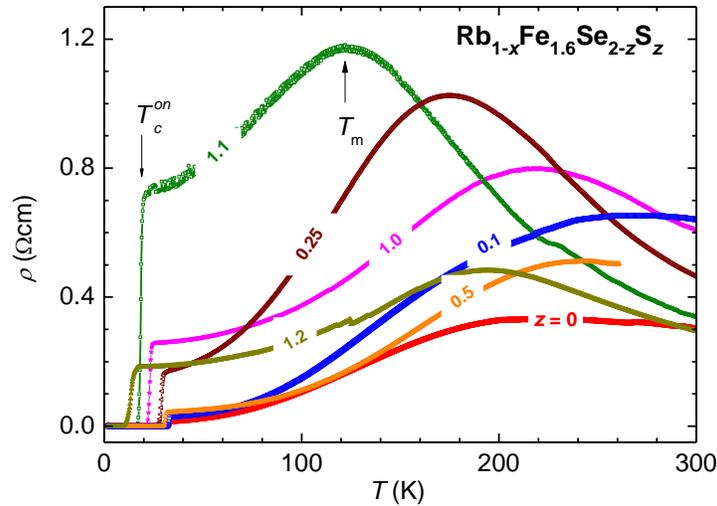

FIG. 12. Temperature dependent resistivity for superconducting $Rb_{1-x}Fe_{2-y}Se_{2-z}S_z$.

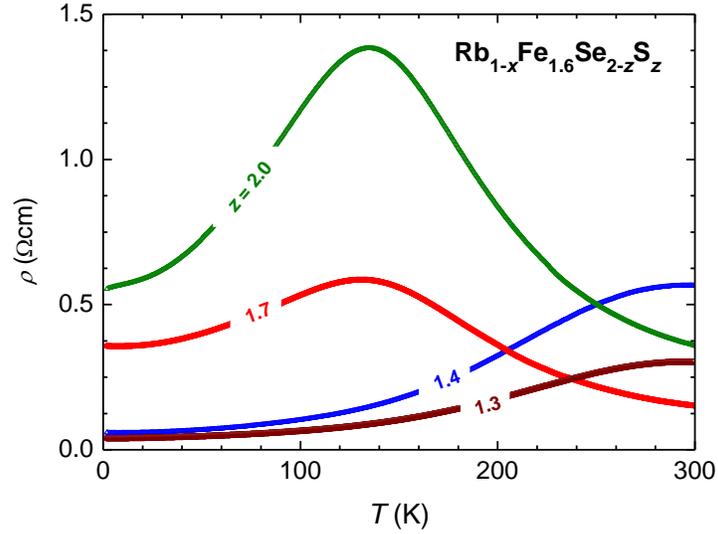

FIG. 13. Temperature dependent resistivity for non-superconducting $Rb_{1-x}Fe_{2-y}Se_{2-z}S_z$.

Fig. 14 demonstrates the temperature dependent resistivity measured in different magnetic fields applied in the vicinity of the superconducting transition for $Rb_{1-x}Fe_{2-y}Se_{2-z}S_z$ with substitution $z$ = 0.1, 0.25, 1.0, and 1.2. The data for other substitutions are given in the Supplemental Materials (Fig. 6SM). In zero field, the transition temperature of the substituted samples determined at the level of a 90% drop of the normal-state resistivity differs by 1 to 2 K from that of the onset temperature $T_c$ estimated from the susceptibility measurements. This indicates increasing inhomogeneity compared to pure $Rb_{0.8}Fe_{1.6}Se_2$ in which this difference does not exceed 0.1K [14]. When increasing the magnetic field, the resistivity curves are shifted to lower temperatures. Fig. 15 shows the temperature dependence of the upper critical field $H_{c2}(T)$ for samples with different substitution level estimated by using the criterion of a 90% drop of the normal-state resistivity. The estimation of the upper critical field $H_{c2}(0)$ for $T$ = 0 K was performed with the Werthamer-Helfand-Hohenberg model [34] using the relation $H_{c2}(0)$ = $-0.69T_c(dH_{c2}(T)/dT)|_{T_c}$. The upper critical field increases from 22 T for $z$ = 0 to 35 T with increasing sulfur substitution up to $z$ = 0.25, but then decreases with a further growth of the sulfur content in the samples, going down to the value of 9 T for $z$ = 1.1 (as shown in Fig. 16). It is noteworthy that the sample with $z$ = 0.25 with a smaller $T_c$ than that with $z$ = 0 shows the highest value of the upper critical field.

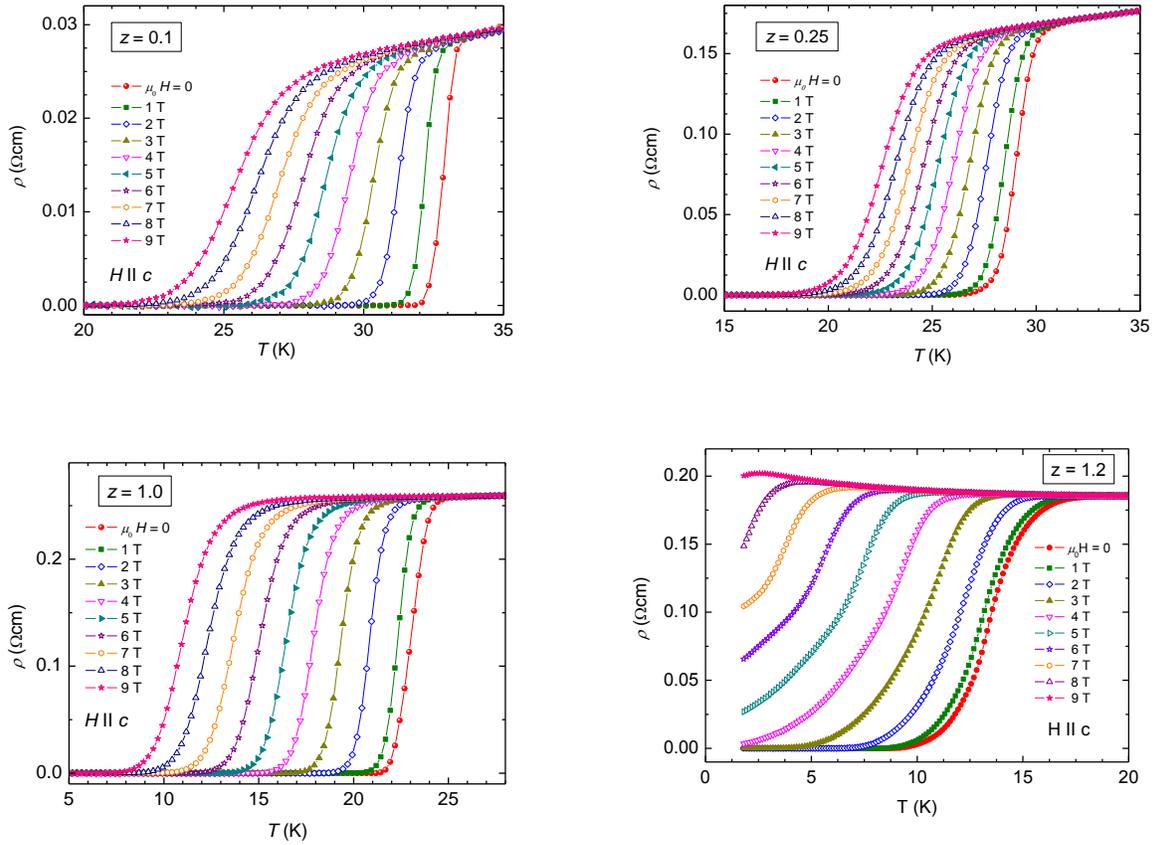

FIG. 14. Temperature dependent resistivity in different applied magnetic fields in the vicinity of superconducting transition for $Rb_{1-x}Fe_{2-y}Se_{2-z}S_z$ with $z = 0.1$, 0.25, 1.0, and 1.2.

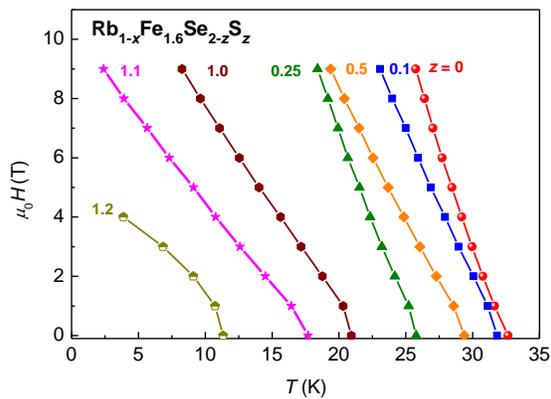

FIG. 15. Temperature dependence of upper critical field $H_{c2}$ for samples with different substitutions.

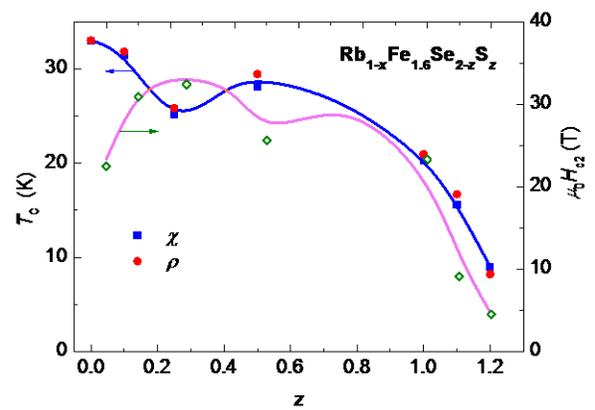

FIG. 16. Concentration dependence of the critical temperature $T_c$ (left scale) and of upper critical field $H_{c2}(0)$ (right scale). Closed circles and squares show $T_c$ estimated from resistivity

and susceptibility measurements, respectively.

## F. Specific heat

Fig. 17 presents the temperature dependent specific heat $C$ for selected samples with different substitutions. In the measured temperature range, the specific heat for both superconducting and non-superconducting samples exhibits quite similar behavior being dominated by the lattice contribution. For the superconducting samples, the anomaly at the critical temperature is hardly detectable in the raw data. It becomes clearly visible only after subtraction of the lattice and magnetic contributions from the total specific heat $C$. The data for the electronic specific heat $C_{el}$ of several non-substituted samples $Rb_{0.8}Fe_{1.6}Se_2$, are shown in Fig. 18.

An important problem for the calculation of $C_{el}$ is related to determination of the phonon, $C_{lat}$, and magnetic, $C_m$, contributions. In Ref. [14] it was found that an insulating sample $Rb_{0.75}Fe_{1.5}Se_2$ exhibits very similar magnetic properties like superconducting $Rb_{0.8}Fe_{1.6}Se_2$. The present study of $Rb_{1-x}Fe_{2-y}Se_{2-z}S_z$ system also revealed quite similar antiferromagnetic behavior of all samples independent on substitution. Therefore, for modelling the phonon and magnetic contributions, the specific heat data for the non-superconducting samples $Rb_{0.75}Fe_{1.5}Se_2$ with $C_{lat}(0)$ for $z = 0$, and $Rb_{0.8}Fe_{1.6}S_2$ with $C_{lat}(2)$ for $z = 2$ were used. For the substituted samples, this contribution was calculated taking into account the respective weight of $C_{lat}(0)$ and $C_{lat}(2)$, e.g., $0.5[C_{lat}(0) + C_{lat}(2)]$ for $C_{lat}(1)$, $0.5[C_{lat}(0) + C_{lat}(1)]$ for $C_{lat}(0.5)]$, *etc*.

The inset in Fig. 17 shows the temperature dependent specific heat in the representation $C/T$ *vs.* $T^2$ at temperatures below 10 K for samples with $z = 0$, 1, and 2. These dependencies display two linear regimes: one below 4.5 K and the other in the temperature range from 7 to 10 K. Assuming that in the lower linear regime the superconducting contribution to heat capacity is much smaller than in the second one, the experimental data at temperatures below 4 K were fitted by the expression $C/T = \gamma + \beta T^2$. Here $\gamma$ is the coefficient for the term in the specific heat that is linear in temperature and the prefactor $\beta$ characterizes the lattice and magnon contributions to the specific heat, which are both proportional to $T^3$ and cannot be estimated independently because the AFM transition temperature $T_N$ and the Debye temperature $\theta_D$ are comparable. The calculated values of the parameters $\gamma$ and $\beta$ are given in Table 3. The phonon and magnetic contributions for the superconducting samples with substitutions $z \leq 1.2$ were corrected for the difference in their effective Debye temperatures, when comparing with the value $\theta_D$ for the modelled non-superconducting contribution [35]. The effective Debye temperature was calculated from the relation $\theta_D = [12\pi^4 k_B N_A Z/(5\beta)]^{1/3}$, where $k_B$ and $N_A$ are the Boltzmann and the Avogadro constants, respectively. Z = 5, is the number of atoms in the unit cell. The calculated values of $\theta_D$ for all studied samples are also presented in Table 3.

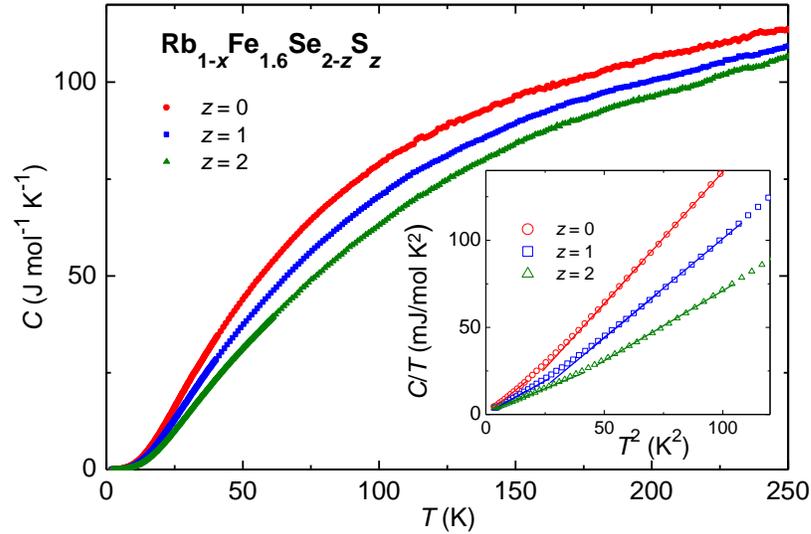

FIG. 17. Temperature dependent specific heat $C$ for selected $Rb_{1-x}Fe_{2-y}Se_{2-z}S_z$ samples with different substitution.

The experimentally determined values of $\gamma$ were found to vary in a range 0.08 - 0.3 mJ/(mol K$^2$) for all samples with substitution $z \leq 1.1$ (with exception of sample with $z = 0.25$) suggesting a low amount of impurities. A similar value of $\gamma = 0.394$ mJ/(mol K$^2$) was reported for a high-quality superconducting sample $K_xFe_{2-y}Se_2$ in Ref. 36. For our samples with $z \geq 1.2$, the value of $\gamma$ increased significantly indicating an increasing amount of the metallic phase. The value of the prefactor $\beta$ shows a continuous decrease with increasing substitution as one naturally anticipates in case of a respective decrease of the molar mass of the samples. However, about 10% difference in the value of $\beta$ for non-substituted samples (with $z = 0$) from the different batches was found, which cannot be fully accounted for by the difference in their compositions. It also should be noticed that the values of $\beta$ for the studied $Rb_{0.8}Fe_{1.6}Se_2$ samples in the course of this work are quite similar to $\beta = 1.018$ mJ/(mol K$^4$) given in Ref. 36 for $K_xFe_{2-y}Se_2$ suggesting a comparable quality of our samples.

The value of the Sommerfeld coefficient in the normal state $\gamma_n$ for superconducting samples was calculated from the temperature dependence of the electronic specific heat using the constraint of entropy conservation at the onset of $T_c$, *i.e.*,

$$\int\limits_0^{T_c} \frac{C_{el}}{T} dT = \int\limits_0^{T_c} \gamma_n dT$$

The values of $\gamma_n$ are also given in Table 3. For non-substituted $Rb_{0.8}Fe_{1.6}Se_2$, the values of $\gamma_n$ differ significantly for the samples from different batches and even from the same batch. To understand the reason of this variation, the specific heat data measured under applied magnetic fields, in which the phonon and magnetic contribution are expected to be identical to those for zero field [36], were analyzed. In Fig. 19, the difference in the $C$ values measured in zero field and a field of 9 T *vs.* temperature is shown for several samples $Rb_{0.8}Fe_{1.6}Se_2$. For all these samples, the λ anomaly at around $T_c$ is quite sharp and its width does not exceed 4 K, being much lower than the shift of $T_c$ by the field of 9 T [14,36]. Importantly, the amplitude of the anomaly at $T_c$ is very similar for different samples indicating that they exhibit close values of the electronic specific heat $C_{el}$. This indicates that the procedure applied to estimate $C_{el}$ can create significant errors in calculation of the Sommerfeld coefficient $\gamma_n$. Inspection of Fig. 19 reveals the following additional features in the heat-capacity data in zero field: a pronounced tail at temperatures above $T_c$ indicating fluctuating superconductivity and a step at temperatures between 20 and 27 K, which suggests the presence of additional density of states besides the percolating superconducting ones. It is worth mentioning that even in samples that do not show a superconducting ground state, a broad anomaly in the specific-heat difference $C_{0T}$-$C_{9T}$ in the temperature range from 20 K to 40 K was observed. We assume that it can be related to non-percolated SC states due to intrinsic inhomogeneities of the samples. Evidently, these features cannot be accounted for by the modelled lattice and magnetic contribution.

Under assumption that the values of $C_{el}$ for different samples with $z = 0$ are the same and in order to minimize the errors when subtracting phonon and magnon contributions, we averaged the experimentally determined specific heat data over seven measured samples. The calculated values of the parameters for the average data are given in Table 3. The value of the Sommerfeld coefficient $\gamma_n$ for the averaged data equals 10.3 mJ/(mol K$^2$). The reduced specific heat jump at $T_c$, $\Delta C/\gamma_n T_c$, for the averaged data was 0.79. For the sample with the lowest calculated value of $\gamma_n$ $_=6.2$ mJ/(mol $K^2$), the reduced specific jump at $T_c$ was 1.31, which is slightly lower than the BCS estimate of 1.43 for the weak-coupling limit. It differs from the value of $\Delta C/\gamma_n T_c = 1.93$ obtained for $K_x Fe_{2-y} Se_2$ in Ref. 36, which is characteristic for strong coupling. The reason of this significant difference between two related systems needs additional study. To clarify this problem, an independent method of evaluation of $\gamma_n$ is desired.

Table 3. Parameters calculated from the specific heat data for $Rb_{1-x}Fe_{2-y}Se_{2-z}S_z$ samples.

| Sample label | Substitution $z$ | $\gamma$ (0K) mJ/(mol $K^2$) | $\beta$ mJ/(mol $K^4$) | $\theta_D$ K | $\gamma_n$ mJ/(mol $K^2$) | $N$ States/(eV f.u.) |
|---|---|---|---|---|---|---|
| BR19 NSC | 0 | - | 0.97(1) | 215.4 | - | - |
| BR16s1 | 0 | 0.18 | 1.10(1) | 206.5 | 10.7 | 4.5 |
| BR16s8 | 0 | 0.24 | 1.07(1) | 208.4 | 10.7 | 4.5 |
| BR26s1 | 0 | 0.09 | 0.99(1) | 213.9 | 6.2 | 2.6 |
| BR26s6 | 0 | 0.08 | 1.04(1) | 210.4 | 7.1 | 3.0 |
| BR26s11 | 0 | 0.30 | 1.04(1) | 210.4 | 13.7 | 5.8 |
| BR26s12 | 0 | 0.10 | 1.02(1) | 211.8 | 9.8 | 4.2 |
| BR26s13 | 0 | 0.09 | 0.98(1) | 214.6 | 7.4 | 3.1 |
| *Average* | *0* | *0.15* | *1.02(1)* | *211.8* | *10.3* | *4.4* |
| BR98 | 0.25 | 0.51 | 0.91(2) | 220.0 | 5.3 | 2.3 |
| BR96 | 0.5 | 0.23 | 0.83(1) | 226.9 | 3.9 | 1.7 |
| BR80 | 1.0 | 0.12 | 0.79(1) | 230.6 | 3.5 | 1.5 |
| BR87 | 1.0 | 0.25 | 0.78(1) | 231.6 | 3.7 | 1.6 |
| BR82 | 1.1 | 0.23 | 0.77(1) | 232.6 | 1.5 | 0.6 |
| BR107 | 1.2 | 0.81 | 0.764(6) | 233.2 | 1.4 | 0.6 |
| BR109 | 1.3 | 0.81 | 0.764(6) | 233.2 | | |
| BR97 | 2.0 | 1.51 | 0.533(6) | 262.9 | - | |

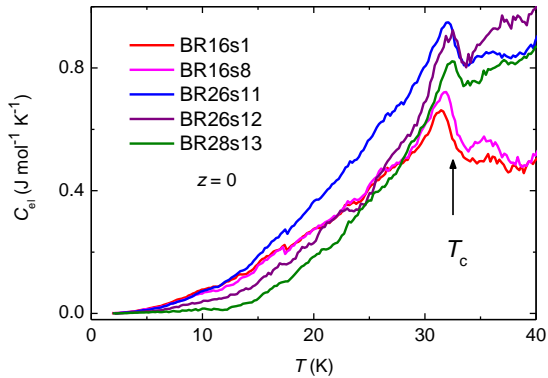

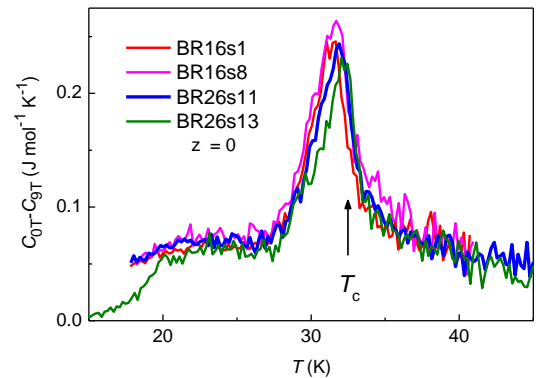

FIG. 18. Temperature dependent electronic specific heat $C_{el}$ for $Rb_{0.8}Fe_{1.6}Se_2$ samples ($z = 0$) from different batches. Arrow marks $T_c$ taken from the susceptibility data.

FIG. 19. Difference of the heat-capacity values measured in zero field and in an external magnetic field of 9 T $vs.$ temperature $T$ for several samples of non-substituted $Rb_{0.8}Fe_{1.6}Se_2$ ($z = 0$).

Fig. 20 shows the electronic specific heat $C_{el}$ $vs.$ temperature normalized to the critical temperature $T_c$ for superconducting $Rb_{1-x}Fe_{2-y}Se_{2-z}S_z$ samples with different substitutions. The magnitude of the $\lambda$ anomaly at $T_c$ shows a continuous decrease with substitution suggesting a reduction of the amount of the superconducting phase. With increasing substitution, a reduction of the values of $C_{el}$ and of the Sommerfeld coefficient $\gamma_n$ takes place. This indicates that the suppression of the superconductivity with increasing substitution is accompanied by the reduction of the density of states at the Fermi energy. This conclusion is further supported by the data presented in Fig. 21, which shows the difference in the experimental specific heat $C$ measured in zero field and in a field of 9 T $vs.$ temperature for $Rb_{1-x}Fe_{2-y}Se_{2-z}S_z$ samples with different substitutions. With increasing $z$ from 0 to 1.0, a significant reduction, of the magnitude of the $\lambda$ anomaly in the total specific heat $C$ at $T_c$ estimated from the difference $C_{0T}$-$C_{9T}$, by a factor of 6 occurs. A somewhat smaller reduction, by a factor of 3 to 4, was calculated from the respective change of $\gamma_n$ for these samples (see Table 3). A more precise quantitative estimate of the reduction of the density of states with substitution seems to be difficult because of the uncertainty in calculating the electronic specific heat mentioned above and because of the poor statistics for samples with $z > 0$. It is important to mention here that the reduction of the density of states at the Fermi energy with substitution, derived from the specific heat data is in good agreement with the results of Ref. 24, where it was found that with increasing substitution of S for Se, the orbital-selective Mott transition shifts to higher temperatures due to reduction of correlations in the $d_{xy}$ channel. We therefore attribute the observed suppression of $T_c$ in the $Rb_{1-x}Fe_{2-y}Se_{2-z}S_z$ system to this mechanism.

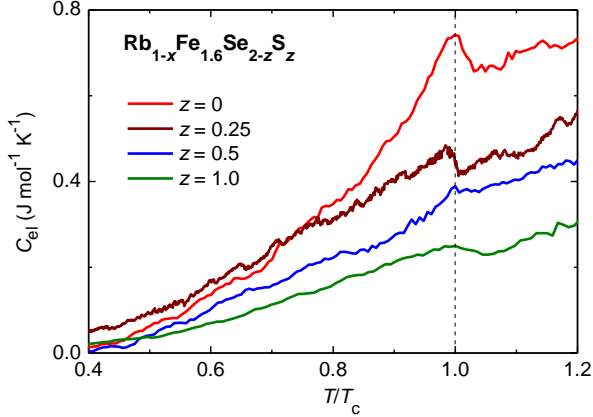
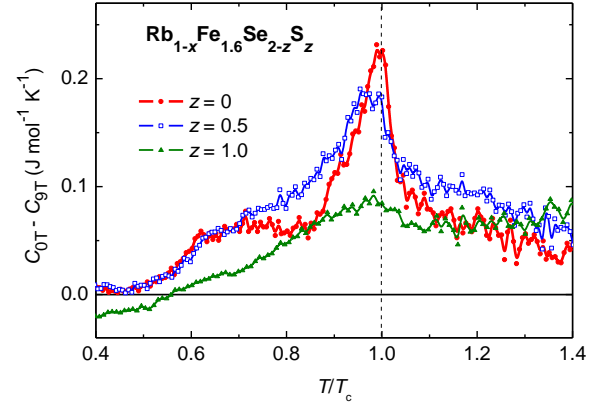

FIG. 20. Temperature dependent electronic specific heat $C_{el}$ for $Rb_{1-x}Fe_{2-y}Se_{2-z}S_z$ samples with different substitution $z$.

FIG. 21. Difference in the heat-capacity values measured in zero field and in field of 9 T *vs.* temperature $T$ for $Rb_{1-x}Fe_{2-y}Se_{2-z}S_z$ samples with different substitution $z$.

## G. Phase diagram and conclusions

Fig. 22 presents the phase diagram of $Rb_{1-x}Fe_{2-y}Se_{2-z}S_z$, which summarizes the results of our studies. At the lowest temperatures, the ground state of the samples with substitution $z \leq 1.2$ is superconducting coexisting with the AFM state. With increasing substitution, a reduction of the SC transition temperature $T_c$ takes place. The percolation threshold for the SC state is within the range of concentrations 1.2 and 1.3. The AFM state is present in all samples independent on the substitution level. The AFM phase has a Fe-vacancy ordered structure below the structural transition at $T_s$. The transition temperature into the AFM state $T_N$ shows a monotonous decrease indicating a weakening of AFM interactions with increasing substitution of Se by S ions. Since the AFM correlations are important for triggering the SC state via proximity effect, the weakening of the AFM interactions can contribute to the observed suppression of the superconductivity in this system. This scenario is supported by the fact that the temperature of the AFM ordering $T_N$ and the temperature of appearance of the minority SC phase $T_p$ cross in the same concentration region where the SC temperature $T_c$ goes to zero. This is reminiscent also of an external pressure experiments on the SC $Rb_{0.8}Fe_{1.6}Se_2$, which revealed that suppression of the SC phase takes place concomitant with suppression of the AFM phase [37]. A similar effect takes place in $Rb_{1-x}Fe_{2-y}Se_{2-z}S_z$ system due to an increase of the chemical pressure with substitution of Se by S ions.

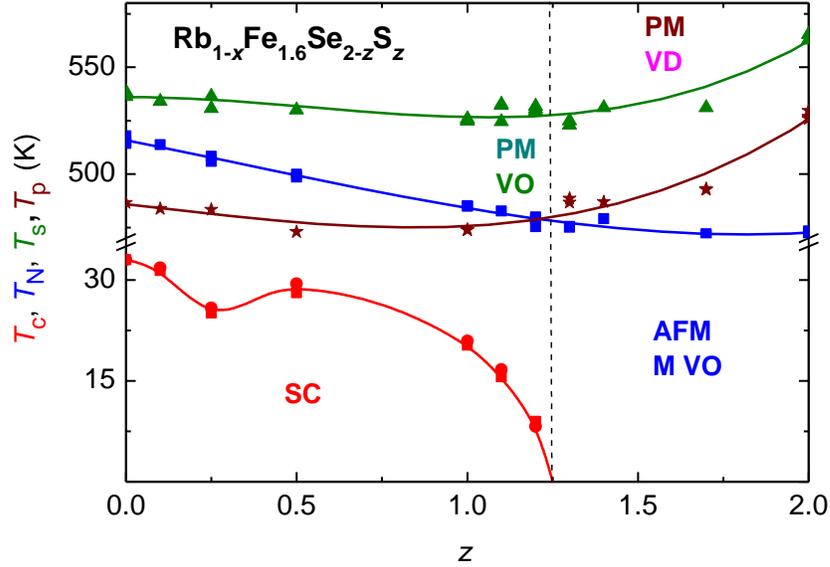

FIG. 22. *T-z* phase diagram of the $Rb_{1-x}Fe_{2-y}Se_{2-z}S_z$ system. SC - superconducting state, AFM M VO - antiferromagnetic metallic vacancy ordered, PM VO - paramagnetic vacancy ordered, PM VD - paramagnetic vacancy disordered. Vertical dashed line separates SC and non-superconducting samples.

In conclusion, our detailed structural, magnetic, conductivity, and thermodynamic studies of $Rb_{1-x}Fe_{2-y}Se_{2-z}S_z$ revealed several important peculiarities of this system:

1. The superconducting state exists up to $z = 1.2$. With increasing $z$, the temperature of the SC transition $T_c$ shows a non-monotonic drop from 32.4 K for $z = 0$ to 10 K for $z = 1.2$. Similar non-monotonic behavior with substitution exhibits the second critical field $H_{c2}(0)$, which reaches a value of 35 T for a substitution $z = 0.25$.

2. The anisotropic AFM state is a characteristic feature for all samples of $Rb_{1-x}Fe_{2-y}Se_{2-z}S_z$ independent of substitution. With increasing substitution, the transition temperature into the AFM state $T_N$ shows a continuous decrease from 515 K for $z = 0$ to 472 K for $z = 2$.

3. The Fe-vacancy ordered crystal structure of the studied samples exists within the entire range of substitution. The temperature of the structural transformation into the vacancy-ordered state $T_s$ changes non-monotonously with substitution. It decreases from 538 K ($z = 0$) to 523 K for $z = 1.3$ and then increases again to 563 K for $z = 2$.

4. The observed variations of the bond distances and bond angles in the Fe tetrahedrons indicate a decrease of the structural distortions with substitution.

5. The SC and AFM phases are coexisting in a phase-separated arrangement. For non-substituted samples ($z = 0$), the SC stripes are of μm size. Their composition corresponds to $Rb_{0.705(25)}Fe_{2.017(10)}Se_2$, while that of the AFM matrix corresponds to $Rb_{0.8}Fe_{1.6}Se_2$. For the substituted samples, the phase separation of the SC and AFM phases is obviously realized on lower length scales.

6. Above the SC transition and below 140 K, the samples with substitutions $z \leq 1.2$ manifest a metallic-like conductivity, while at higher temperatures, a metal-to-semiconductor transition takes place. The ground state of the samples with higher substitution, including $z = 2$, is also metallic.

7. A significant reduction of the λ anomaly in the specific heat at the SC transition with increasing substitution indicates a reduction of the density of states at the Fermi energy that can account for the observed suppression of the superconducting state.

**Acknowledgements.**

This work was supported by the Transregional Research Collaboration TRR 80 (Augsburg, Munich, and Stuttgart) and by Institutional project 15.817.02.06F (Moldova). The support of the Grant for Young Researchers CSSDT 18.80012.02.10F (Moldova) is also acknowledged.

**References**

1. E. Dagotto, The unexpected properties of alkali metal ion selenide superconductors, Rev. Mod. Phys. **85**, 849 (2013).

2. Wei Bao, Structure, magnetic order and excitations in the 245 family of Fe-based superconductors, J. Phys. Condens. Matter **27**, 023201 (2015).

3. A. Krzton-Maziopa, V. Svitlyk, E. Pomjakushina, R. Puzniak, and K. Conder, Superconductivity in alkali metal intercalated iron selenides, J. Phys.: Condens. Matter **28**, 0293002 (2016).

4. W. Bao, Q. Huang, G.F. Chen, M.A. Green, D.M. Wang, J.B. He, X.Q. Wang, and Y. Qiu, A novel large moment antiferromagnetic order in $K_{0.8}Fe_{1.6}Se_2$ superconductor, Chin. Phys. Lett. **28**, 086104 (2011).

5. P. Dudin, D. Herriott, T. Davies, A. Krzton-Maziopa, E. Pomjakushina, K. Conder, C. Cacho, J.R. Yatea, and S.C. Speller, Imaging the local electronic and magnetic properties of intrinsically phase separated $Rb_xFe_{2-y}Se_2$ superconductor using scanning microscopy techniques, Supercond. Sci. Technol. **32**, 044005 (2019).

6. J. Guo, S. Jin, G. Wang, S. Wang, K. Yhu, T. Zhou, M. He, and X. Chen, Superconductivity in the iron selenide $K_xFe_2Se_2$ ($0 \leq x \leq 1.0$), Phys. Rev. B **82**, 180520 (2010).

7. A. Krzton-Maziopa, Z. Shermadini, E. Pomjakushina, V. Pomjakushin, M. Bendele, A. Amato, R. Khasanov, H. Luetkens, and K. Conder, Synthesis and crystal growth of $Cs_{0.8}(FeSe_{0.98})_2$: a new iron-based superconductor with $T_c = 27$ K, J. Phys. Condens. Matter **23**, 052203 (2011).

8. J.J. Ying, X.F.Wang, X.G. Luo, A.F.Wang, M. Zhang, Y.J. Yan, Z.J. Xiang, R.H. Liu, P. Cheng, G.J. Ye, and X.H. Chen, Superconductivity and magnetic properties of single crystals of $K_{0.75}Fe_{1.66}Se_2$ and $Cs_{0.81}Fe_{1.61}Se_2$, Phys. Rev. B **83**, 212502 (2011).

9. C.H. Li, B. Shen, F. Han, X. Zhu, and H.H. Wen, Transport properties and anisotropy of $Rb_{1-x}Fe_{2-y}Se_2$ single crystals, Phys. Rev. B **83**, 184521 (2011).

10. X.G. Luo, X.F. Wang, J.J. Ying, Y.J. Yan, Z.Y. Li, M. Zhang, A F. Wang, P. Cheng, Z.J. Xiang, G.J. Ye, R.H. Liu, and X.H. Chen, Crystal structure, physical properties and superconductivity in $A_xFe_2Se_2$ single crystals, New J. Phys. **13**, 053011 (2011).

11. Y. Mizuguchi, H. Takeya, Y. Kawasaki, T. Ozaki, S. Tsuda, T. Yamaguchi, and Y. Takano, Transport properties of the new Fe-based superconductor $K_xFe_2Se_2$ ($T_c = 33$ K), Appl. Phys. Lett. **98**, 042511 (2011).

12. A.F. Wang, J.J. Ying, Y.J. Yan, R.H. Liu, X.G. Luo, Z.Y. Li, X.F. Wang, M. Zhang, G.J. Ye, P. Cheng, Z.J. Xiang, and X.H. Chen, Superconductivity at 32 K in single-crystalline $Rb_xFe_{2-y}Se_2$, Phys. Rev. B **83**, 060512 (2011).

13. H.D.Wang, C.H. Dong, Z.J. Li, Q.H. Mao, S.S. Zhu, C.M. Feng, H.Q. Yuan, and M.H. Fang, Superconductivity at 32 K and anisotropy in $Tl_{0.58}Rb_{0.42}Fe_{1.72}Se_2$ crystals, Europhys. Lett. **93**, 47004 (2011).

14. V. Tsurkan, J. Deisenhofer, A. Günther, H.-A. Krug von Nidda, S. Widmann, and A. Loidl, Anisotropic magnetism, superconductivity, and the phase diagram of $Rb_{1-x}Fe_{2-y}Se_2$, Phys. Rev. B **84**, 144520 (2011).

15. S. Medvedev, T.M. McQueen, I.A. Troyan, T. Palasyuk, M.I. Eremets, R.J. Cava, S. Naghavi, F. Casper, V. Ksenofontov, G. Wortmann, and C. Felser, Electronic and magnetic phase diagram of $\beta$-$Fe_{1.01}Se$ with superconductivity at 36.7 K under pressure, Nat. Mater. **8**, 630 (2009).

16. S. Margadonna, Y. Takabayashi, Y. Ohishi, Y. Mizuguchi, Y. Takano, T. Kagayama, T. Nakagawa, M. Takata, and K. Prassides, Pressure evolution of the low-temperature crystal structure and bonding of the superconductor FeSe ($T_c$=37 K), Phys. Rev. B **80**, 064506 (2009).

17. F.-C. Hsu, J.Y. Luo, K.W. Yeh, T.K. Chen, T.W. Huang, P.M. Wu, Y.-C. Lee, Y.L. Huang, Y.-Y. Chu, D.C. Yan, and M.-K. Wu, Superconductivity in the PbO-type structure α-FeSe, Proc. Natl. Acad. Sci. U.S.A. **105**, 14262 (2008).

18. S. He, J. He, W. Zhang, L. Zhao, D. Liu, X. Liu, D. Mou, Y.-B. Ou, Q.-Y. Wang, Z. Li, L. Wang, Y. Peng,Y. Liu, C. Chen, L. Yu, G. Liu, X. Dong, J. Zhang, C. Chen, Z. Xu, X. Chen, X. Ma, Q. Xue and X. J. Zhou, Phase diagram and electronic indication of high-temperature superconductivity at 65 K in single-layer FeSe films, Nat. Mater. **12**, 605 (2013).

19. J.F. Ge, Z.L. Liu, C.H. Liu, C.L. Gao, D. Qian, Q.K. Xue, Y. Liu, and J.F. Jia, Superconductivity above 100 K in single-layer FeSe films on doped $SrTiO_3$, Nat. Mater. **14**, 285 (2015).

20. M.H. Fang, H.M. Pham, B. Qian, T.J. Liu, E.K. Vehstedt, Y. Liu, L. Spinu, and Z.Q. Mao, Superconductivity close to magnetic instability in $Fe(Se_{1-x}Te_x)_{0.82}$, Phys. Rev. B **78**, 224503 (2008).

21. K.W. Yeh, T.-W. Huang, Y. Huang, T.-K. Chen, F.-C. Hsu, P.M. Wu, Y.-C. Lee, Y.-Y. Chu, C.-L. Chen, J.-Y. Luo, D.-C. Yan, and M.-K. Wu, Tellurium substitution effect on superconductivity of the α-phase iron selenide, Europhys. Lett. **84**, 37002 (2008).

22. H. Lei, M. Abeykoon, E.S. Bojin, K. Wang, J.B. Warren, and C. Petrovic, Phase diagram of $K_xFe_{2-y}Se_{2-z}S_z$ and the suppression of its superconducting state by an Fe2−Se/S tetrahedron distortion, Phys. Rev. Lett. **107**, 137002 (2011).

23. M. Yi, D.H. Lu, R.Yu, S.C. Riggs, J.-H. Chu, B. Lv, Z.K. Liu, M. Lu, Y.-T. Cui, M. Hashimoto, S.-K. Mo, Z. Hussain, C.W. Chu, I.R. Fisher, Q. Si, and Z.-X. Shen, Observation of temperature-induced crossover to an orbital-selective Mott phase in $A_xFe_{2-y}Se_2$ (A = K, Rb) Superconductors, Phys. Rev. Lett. **110**, 067003 (2013).

24. Zhe Wang, V. Tsurkan, M. Schmidt, A. Loidl, and J. Deisenhofer, Tuning orbital-selective correlations in superconducting $Rb_{0.75}$ $Fe_{1.6}$ $Se_{2-z}S_z$, Phys. Rev. B **93**, 104522 (2016).

25. Zhe Wang, M. Schmidt, J. Fischer, V. Tsurkan, M. Greger, D. Vollhardt, A. Loidl and J. Deisenhofer, Orbital-selective metal–insulator transition and gap formation above $T_c$ in superconducting $Rb_{1-x}Fe_{2-y}Se_2$, Nature Communications **5**, 3202 (2014).

26. G.M. Sheldrick, Crystal structure refinement with *SHELXL*, Acta Cryst. **C71**, 3 (2015).

27. L.J. Farrugia. *WinGX* suite for small-molecule single-crystal crystallography, J. Appl. Crystallogr. **32**, 837 (1999).

28. A. Charnukha, A. Cvitkovic, T. Prokscha, D. Propper, N. Ocelic, A. Suter, Z. Salman, E. Morenzoni, J. Deisenhofer, V. Tsurkan, A. Loidl, B. Keimer, and A.V. Boris, Nanoscale layering of antiferromagnetic and superconducting phases in $Rb_2Fe_4Se_5$ single crystals, Phys. Rev. Lett. **109**, 017003 (2012).

29. V.Yu. Pomjakushin, A. Krzton-Maziopa, E. Pomjakushina, K. Conder, D. Chernyshov, V. Svitlyk, and A. Bosak. Intrinsic phase separation in the antiferromagnetic superconductor $Rb_xFe_{2-y}Se_2$: a diffraction study, J. Phys.: Condens. Matter **24** (435701 (2012).

30. Y. Texier, J. Deisenhofer, V. Tsurkan, A. Loidl, D.S. Inosov, G. Friemel, and J. Bobroff, NMR study in the iron-selenide $Rb_{0.74}Fe_{1.6}Se_2$: Determination of the superconducting phase as iron vacancy-free $Rb_{0.3}Fe_2Se_2$, Phys. Rev. Lett. **108**, 237002 (2012).

31. V. Ksenofontov *et al.*, to be published.

32. P. Zavalij, Wei Bao, X.F. Wang J.J. Ying, X.H. Chen, D.M. Wang, J.B. He, X.Q. Wang, G.F. Chen, P.-Y. Hsieh, Q. Huang, and M.A. Green, Structure of vacancy-ordered single-crystalline superconducting potassium iron selenide. Phys. Rev. B **83**, 132509 (2011).

33. S. Weyeneth, M. Bendele, F. von Rohr, P. Dluzewski, R. Puzniak, A. Krzton-Maziopa, S. Bosma, Z. Guguchia, R. Khasanov, Z. Shermadini, A. Amato, E. Pomjakushina, K. Conder, A. Schilling, and H. Keller, Superconductivity and magnetism in $Rb_xFe_{2-y}Se_2$: Impact of thermal treatment on mesoscopic phase separation, Phys. Rev. B **86**, 134530 (2012).

34. N.R. Werthamer, E. Helfand, and P.C. Hohenberg, Temperature and purity dependence of the superconducting critical field, $H_{c2}$. III. Electron spin and spin-orbit effects, Phys. Rev. **147**, 295 (1966).

35. The Debye temperature is effective in a sense that it contains the magnetic contribution as well. However, since the modelled specific heat also includes a magnetic term, this does not influence the calculations of the electronic specific heat.

36. B. Zeng, B. Shen, G.F. Chen, J.B. He, D.M. Wang, C.H. Li, and H.H. Wen, Nodeless superconductivity of single crystalline $K_xFe_{2-y}Se_2$ revealed by the low-temperature specific heat, Phys. Rev. B **83**, 144511 (2011).

37. V. Ksenofontov, S.A. Medvedev, L.M. Schoop, G. Wortmann, T. Palasyuk, V. Tsurkan, J. Deisenhofer, A. Loidl, C. Felser, Superconductivity and magnetism in $Rb_{0.8}Fe_{1.6}Se_2$ under pressure, Phys. Rev B **85**, 214519 (2012).

# STRUCTURE, SUPERCONDUCTIVITY, AND MAGNETISM IN $Rb_{1-x}Fe_{1.6}Se_{2-z}S_z$

D. Croitori,[1] I. Filippova,[1] V. Kravtsov,[1] A. Günther,[2] S. Widmann,[2] D. Reuter,[2]
H.-A. Krug von Nidda,[2] J. Deisenhofer,[2] A. Loidl,[2] and V. Tsurkan[1.2]

[1]*Institute of Applied Physics, MD-2028 Chisinau, Republic of Moldova*
[2]*Experimental Physics V, Center for Electronic Correlations and Magnetism, University of Augsburg, 86135
Augsburg, Germany*

**Date: 04 September 2019**

## SUPPLEMENTAL MATERIAL

### A. Microstructure of samples

(a)

(b)

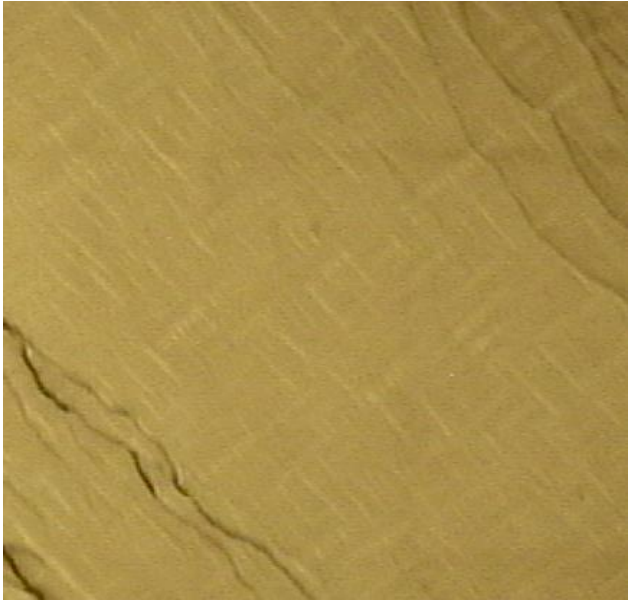
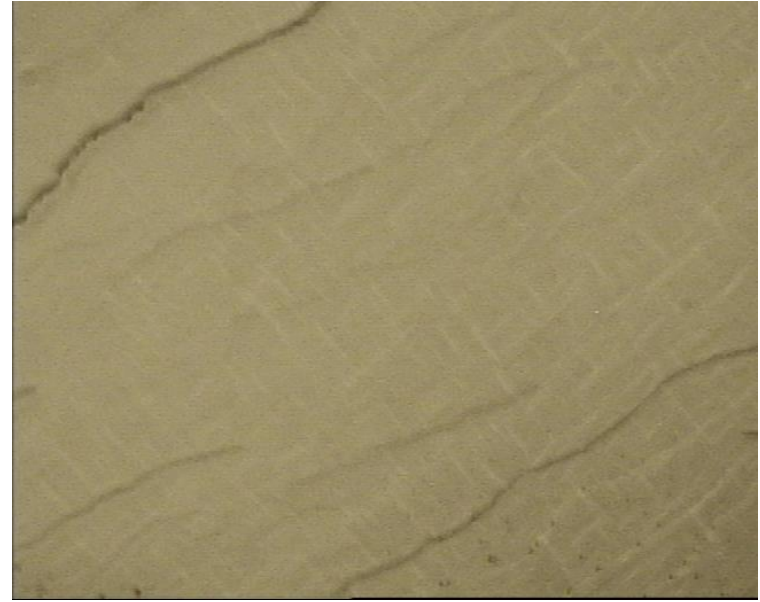

(c) (d)

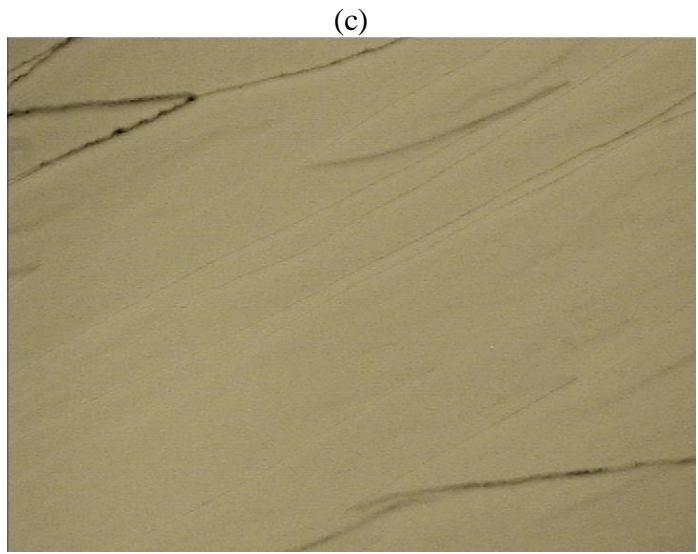 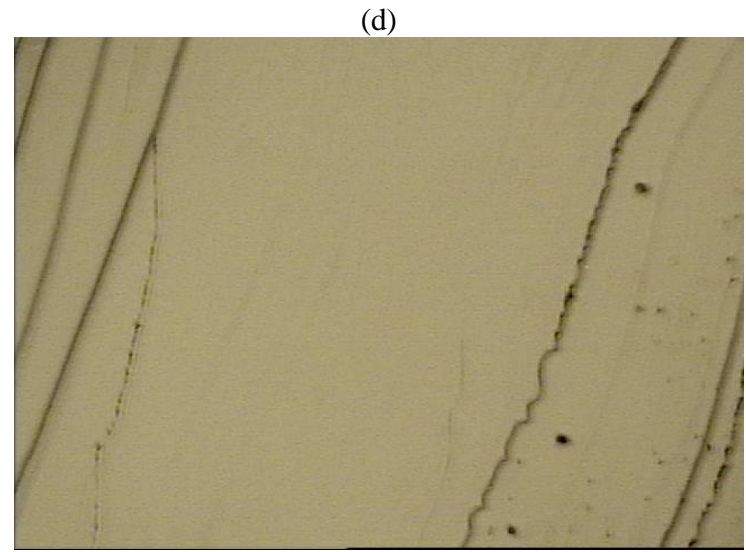

FIG. 1SM. Optical image of surface of $Rb_{1-x}Fe_{2-y}Se_{2-z}S_z$ samples with different substitutions (a) $z = 0$ (sample BR28), (b) $z = 0$ (sample BR16), (c) $z = 0.1$ (BR100), (d) $z = 1.4$ (BR101). All images are taken with the same magnification (x600).

**B. Structural data**

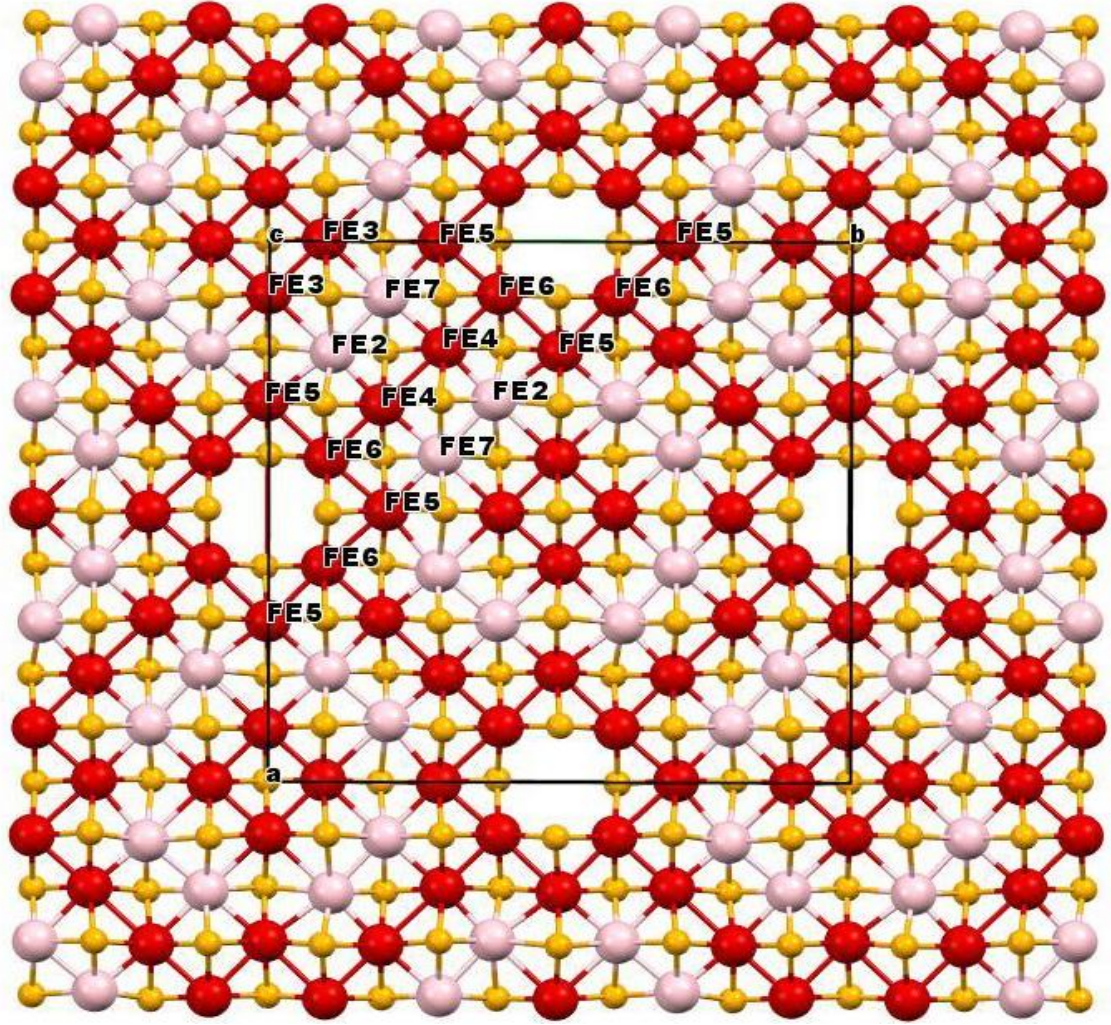

FIG. 2SM. Crystal structure of Rb$_{1-x}$Fe$_{2-y}$Se$_{2-z}$S$_z$ described in space group $I4/m$ within $5 \times 5 \times 1$ supercell.

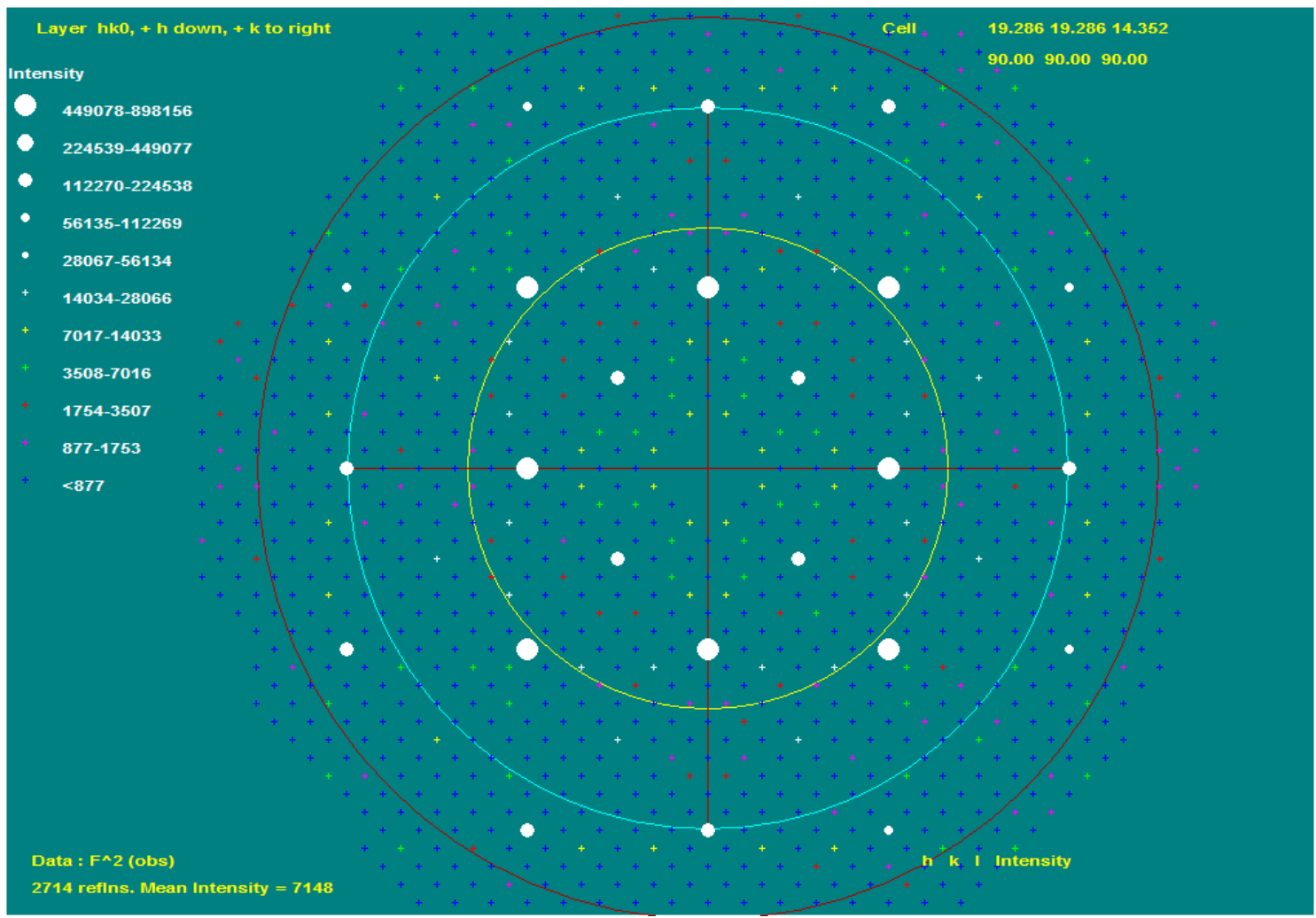

FIG. 3SM (a). Reciprocal lattice plot for Rb$_{1-x}$Fe$_{2-y}$Se$_{2-z}$S$_z$ crystal with $z = 1$ for $5 \times 5 \times 1$ supercell.

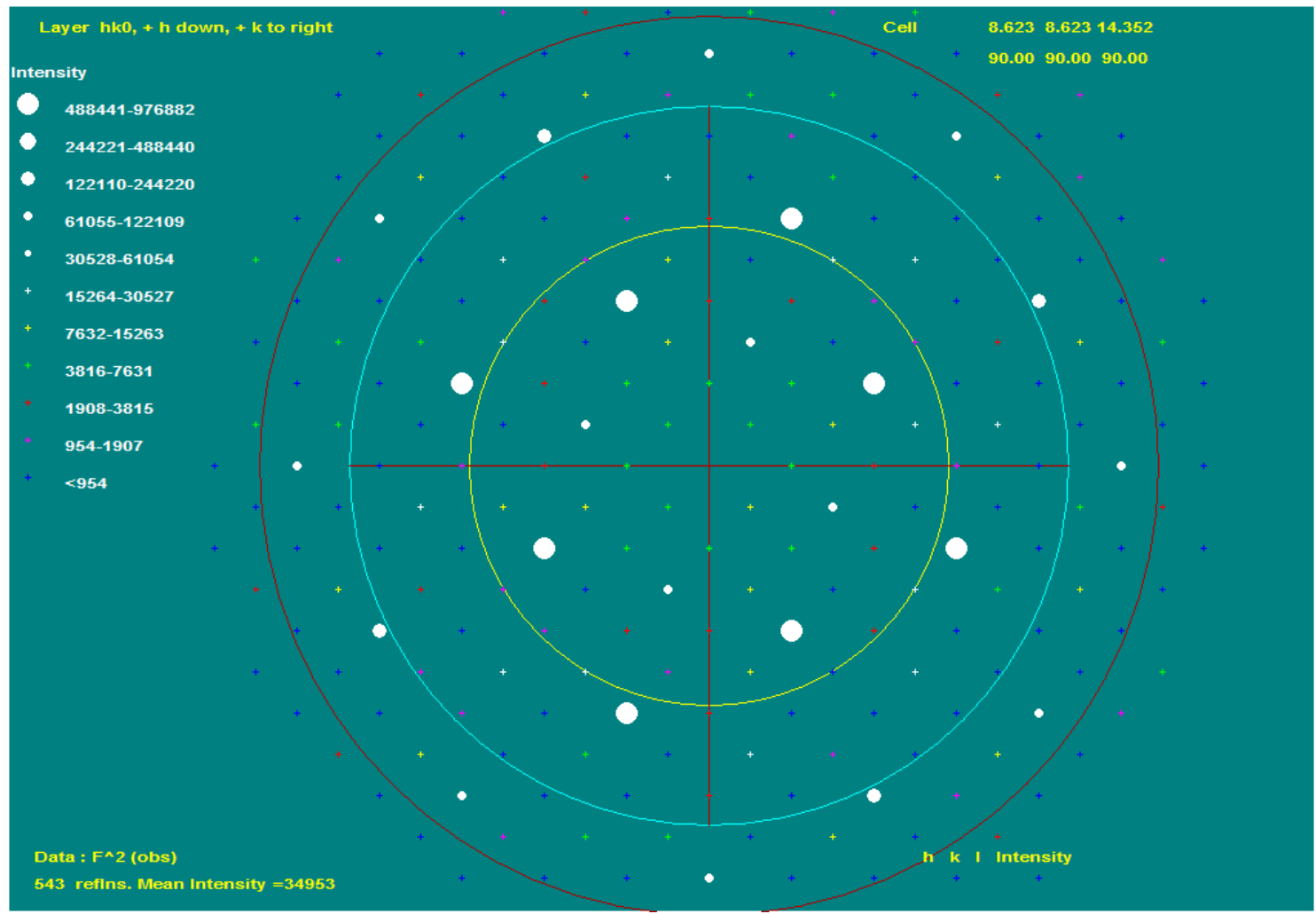

FIG. 3SM (b). Reciprocal lattice plot for Rb$_{1-x}$Fe$_{2-y}$Se$_{2-z}$S$_z$ crystal with $z = 1$ for $\sqrt{5} \times \sqrt{5} \times 1$ cell

Table 1SM. Crystal data and details of structural refinement of Rb$_{1-x}$Fe$_{1.6}$Se$_{2-z}$S$_z$ within 5 × 5 × 1 supercell in space group $I4/m$.

| $z$ (as charged) | 0 | 0.1 | 0.5 | 1.0 | 1.1 | 1.2 | 1.4 | 1.7 | 2 |
|---|---|---|---|---|---|---|---|---|---|
| Formula weight | 314.66 | 299.46 | 271.89 | 254.38 | 251.22 | 248.88 | 236.14 | 222.00 | 218.59 |
| $a = b$ (Å) $c$ (Å) | 19.6765(8) 14.5847(15) | 19.5778(7) 14.5787(10) | 19.4730(6) 14.4762(7) | 19.2864(6) 14.3516(7) | 19.2828(6) 14.3018(9) | 19.2481(8) 14.3077(9) | 19.1242(10) 14.2389(10) | 19.0865(7) 14.1521(7) | 18.9348(8) 14.0386(10) |
| Volume (Å$^3$) | 5646.7(7) | 5587.9(5) | 5489.3(4) | 5338.3(4) | 5317.8(5) | 5300.9(6) | 5207.7(7) | 5155.5(5) | 5033.2(6) |
| Reflections collected / unique | 45872 / 2730 $R_{int}$ = 0.2419 | 34558 / 2704 $R_{int}$ = 0.2413 | 35095 / 2667 $R_{int}$ = 0.1498 | 52362/3352 $R_{int}$ = 0.1757 | 41793 / 3330 $R_{int}$ = 0.1870 | 34994/2564 $R_{int}$ = 0.1924 | 42357 / 2526 $R_{int}$ = 0.1662 | 40724 / 2511 $R_{int}$ = 0.1247 | 36487/2437 $R_{int}$ = 0.1457 |
| $GooF$ | 1.007 | 1.001 | 1.006 | 1.002 | 1.000 | 1.000 | 1.005 | 1.004 | 1.022 |
| $R_1$. $wR_2$ [$I>2\sigma(I)$] | 0.0594. 0.1418 | 0.0873 0.2412 | 0.0910. 0.1833 | 0.0863 0.1673 | 0.0981. 0.1772 | 0.0850 0.1511 | 0.1006. 0.2019 | 0.0702. 0.1434 | 0.0794 0.1634 |
| $R_1$. $wR_2$ (all data) | 0.2098. 0.1774 | 0.1678 0.1967 | 0.2095. 0.2163 | 0.2007 0.2283 | 0.2784. 0.2160 | 0.2461 0.1842 | 0.2187. 0.2357 | 0.1946. 0.1684 | 0.1982 0.1970 |

Table 2SM. Crystal data for Rb$_{1-x}$Fe$_{1.6}$Se$_{2-z}$S$_z$ and details on the structural refinement within √5 × √5 ×1 cell in space group $I4/m$.

| $z$ (as charged) | 0 | 0.1 | 0.5 | 1.0 | 1.1 | 1.2 | 1.4 | 1.7 | 2 |
|---|---|---|---|---|---|---|---|---|---|
| x-ray composition | Rb$_{0.80}$Fe$_{1.61}$ Se$_2$ | Rb$_{0.77}$Fe$_{1.60}$ S$_{0.15}$Se$_{1.85}$ | Rb$_{0.77}$Fe$_{1.60}$ S$_{0.54}$Se$_{1.46}$ | Rb$_{0.76}$Fe$_{1.60}$ SSe | Rb$_{0.87}$Fe$_{1.63}$ S$_{1.1}$Se$_{0.9}$ | Rb$_{0.76}$Fe$_{1.60}$ S$_{1.16}$Se$_{0.84}$ | Rb$_{0.80}$Fe$_{1.61}$ S$_{1.51}$Se$_{0.49}$ | Rb$_{0.77}$Fe$_{1.60}$ S$_{1.69}$Se$_{0.31}$ | Rb$_{0.78}$Fe$_{1.59}$ S$_2$ |
| Formula weight | 315.65 | 306.13 | 287.81 | 264.95 | 271.89 | 257.54 | 245.15 | 233.42 | 219.58 |
| $a$ (Å) $c$ (Å) | 8.805(1) 14.588(1) | 8.754(1) 14.579(1) | 8.706(1) 14.480(1) | 8.623(1) 14.352(1) | 8.624(1) 14.304(1) | 8.608(1) 14.310(1) | 8.545(1) 14.235(1) | 8.535(1) 14.153(1) | 8.462(1) 14.045(2) |
| Volume (Å$^3$) | 1131.08(19) | 1117.20(12) | 1097.57(10) | 1067.08(10) | 1063.82(11) | 1060.29(13) | 1039.31(15) | 1031.01(11) | 1005.8(2) |
| $Z$/ $\rho_{calc}$ (g cm$^{-3}$) | 10/ 4.634 | 10/ 4.550 | 10/ 4.354 | 10/ 4.123 | 10/ 4.244 | 10/ 4.033 | 10/ 3.917 | 10/ 3.760 | 10/ 3.625 |
| $\mu$ (mm$^{-1}$) | 29.579 | 28.496 | 25.964 | 22.833 | 23.541 | 21.701 | 19.752 | 18.009 | 15.945 |
| Crystal size (mm) | 0.20 × 0.15 × 0.01 | 0.30 × 0.20 × 0.02 | 0.35 × 0.30 × 0.02 | 0.30 × 0.20 × 0.02 | 0.15 × 0.10 × 0.02 | 0.30 × 0.12 × 0.04 | 0.30 × 0.30 × 0.03 | 0.25 × 0.20 × 0.20 | 0.35 × 0.25 × 0.02 |
| $\theta$ range for data collection | 3.272 - 28.971 | 3.291 - 25.956 | 3.309 - 25.923 | 3.341 - 24.993 | 2.848 - 24.974 | 3.347 - 24.999 | 3.372 - 24.974 | 3.376 - 27.481 | 3.405 - 25.995 |

| (°) | | | | | | | | | |
|---|---|---|---|---|---|---|---|---|---|
| Reflections collected / unique | 11441 / 779 $R_{int} = 0.1363$ | 6985 / 571 $R_{int} = 0.1527$ | 7329 / 561 $R_{int} = 0.1118$ | 8241 / 495 $R_{int} = 0.0828$ | 7166 / 493 $R_{int} = 0.0674$ | 6634/489 $R_{int} = 0.0830$ | 8036 / 478 $R_{int} = 0.0957$ | 9284 / 617 $R_{int} = 0.0681$ | 7574 / 520 $R_{int} = 0.0859$ |
| Data / parameters | 779 / 36 | 571 / 38 | 561 / 38 | 495 / 38 | 493 / 34 | 489/ 38 | 478 / 35 | 617 / 38 | 520 / 36 |
| GooF | 1.005 | 1.000 | 1.002 | 1.002 | 1.000 | 1.008 | 1.005 | 1.004 | 1.004 |
| $R_1$. $wR_2$ [$I>2\sigma(I)$] | 0.0633, 0.1946 | 0.0635, 0.1688 | 0.0570, 0.1985 | 0.0458, 0.147 | 0.0564, 0.1903 | 0.0453, 0.1198 | 0.0671, 0.2149 | 0.0499, 0.1506 | 0.0488, 0.1496 |
| $R_1$. $wR_2$ (all data) | 0.0986, 0.2221 | 0.0919, 0.1887 | 0.0756, 0.2172 | 0.0585, 0.1523 | 0.0742, 0.2079 | 0.0693, 0.1323 | 0.0848, 0.2387 | 0.0739, 0.1659 | 0.0649, 0.1637 |

Table 3SM. Site occupancy for Fe, Rb, Se, and S atoms in $\sqrt{5} \times \sqrt{5} \times 1$ cell.

| $z$ (as charged) | | 0.0 | 0.1 | 0.5 | 1.0 | 1.1 | 1.4 | 1.7 | 2.0 |
|---|---|---|---|---|---|---|---|---|---|
| Atom | Positions | atom site occupancy | | | | | | | |
| Fe1 | 4d | 0.246(13) | 0.302(8) | 0.303(9) | 0.304(13) | 0.2962(15) | 0.323(18) | 0.299(8) | 0.277(10) |
| Fe2 | 16i | 0.943(11) | 0.925(7) | 0.924(8) | 0.922(14) | 0.942(14) | 0.925(18) | 0.924(9) | 0.917(10) |
| Se1 | 4e | 1 | 0.920(9) | 0.731(9) | 0.496(10) | 0.446(13) | 0.249(13) | 0.152(8) | |
| S1 | 4e | | 0.080(9) | 0.269(9) | 0.504(10) | 0.554(13) | 0.751(13)( | 0.848(8) | 1 |
| Se2 | 16i | 1 | 0.924(8) | 0.732(8) | 0.501(8) | 0.451(10) | 0.239(11) | 0.155(6) | |
| S2 | 16i | | 0.076(8) | 0.268(8) | 0.499(8) | 0.549(10) | 0.761(11) | 0.845(6) | 1 |
| Rb1 | 2b | 0.856(11) | 0.822(9) | 0.828(7) | 0.814(10) | 0.931(11) | 0.895(15) | 0.820(9) | 0.817(12) |
| Rb2 | 8h | 0.781(9) | 0.761(9) | 0.756(7) | 0.742(10) | 0.860(8) | 0.776(13) | 0.754(7) | 0.761(10) |

Table 4SM. Bond distances (Å) and angles (°) for $Rb_{1-x}Fe_{1.6}Se_{2-z}S_z$ in $\sqrt{5} \times \sqrt{5} \times 1$ cell.

| $z$ | 0 | 0.1 | 0.5 | 1.0 | 1.1 | 1.4 | 1.7 | 2 |
|---|---|---|---|---|---|---|---|---|
| Fe1−Se2(S2) | 2.488(1) | 2.476(1) | 2.469(1) | 2.439(1) | 2.441(2) | 2.411(2) | 2.409(1) | 2.383(2) |
| Fe2−Se1(S1) | 2.496(2) | 2.486(2) | 2.475(2) | 2.449(1) | 2.445(2) | 2.413(2) | 2.406(2) | 2.366(2) |
| Fe2−Se2(S2) | 2.447(1) 2.448(1) 2.458(1) | 2.437(1) 2.442(1) 2.446(1) | 2.421(1) 2.428(1) 2.430(1) | 2.388(1) 2.395(1) 2.404(1) | 2.390(2) 2.391(2) 2.407(2) | 2.357(2) 2.358(2) 2.375(2) | 2.345(1) 2.349(1) 2.365(1) | 2.306(2) 2.309(2) 2.319(2) |
| Fe1-Fe2 | 2.778(1) | 2.758(1) | 2.736(1) | 2.699(1) | 2.697(1) | 2.668(1) | 2.660(1) | 2.630(1) |
| Fe2-Fe2 | 2.731(1) 2.908(2) | 2.724(1) 2.881(2) | 2.718(1) 2.863(2) | 2.704(1) 2.832(1) | 2.706(1) 2.835(2) | 2.689(1) 2.801(2) | 2.689(1) 2.803(1) | 2.674(1) 2.778(2) |
| Se2-Fe1-Se2 | 110.35(3) 107.73(6) | 110.59(3) 107.26(5) | 110.56(3) 107.32(6) | 110.38(3) 107.66(6) | 110.34(4) 107.74(8) | 110.05(5) 108.32(11) | 109.78(3) 108.85(6) | 110.30(10) 109.06(5) |
| Se2-Fe2-Se2 | 107.29(5) 107.98(7) 112.75(5) | 107.56(7) 107.78(5) 112.97(5) | 107.56(7) 107.77(5) 113.32(5) | 107.69(5) 107.74(7) 113.74(5) | 107.57(6) 107.64(9) 113.91(7) | 107.41(7) 108.24(11) 113.83(8) | 107.04(5) 108.56(7) 114.18(6) | 106.23(7) 109.68(10) 114.49(8) |
| Se2-Fe2-Se1 | 103.03(6) 112.80(5) 112.81(4) | 103.01(6) 112.66(5) 112.81(5) | 103.07(6) 112.37(4) 112.71(4) | 103.45(7) 111.80(4) 112.37(4) | 103.64(9) 111.72(6) 112.32(6) | 104.31(11) 111.20(7) 111.78(7) | 104.61(7) 110.84(5 111.51(4) | 105.79(11) 110.00(6) 110.47(6) |
| Deviation S1 from Fe2 Fe2 Fe2 plane | 1.581 | 1.571 | 1.559 | 1.530 | 1.522 | 1.486 | 1.474 | 1.421 |
| Deviation S2 from Fe2 Fe2 Fe2 Fe1 plane | 1.460 | 1.462 | 1.453 | 1.433 | 1.433 | 1.405 | 1.395 | 1.354 |

Table 5SM. Ratio of bond angles for Fe1 and Fe2 tetrahedrons to angle of ideal tetrahedron in $\sqrt{5} \times \sqrt{5} \times 1$ cell.

| $z$ | 0 | 0.1 | 0.5 | 1.0 | 1.1 | 1.4 | 1.7 | 2 |
|---|---|---|---|---|---|---|---|---|
| $d_{min}/d_{max}$ for Fe2 | 0.980 | 0.980 | 0.978 | 0.975 | 0.977 | 0.977 | 0.975 | 0.975 |
| $\alpha_{max}/\alpha_{ideal}$ for Fe1 | 1.005 | 1.007 | 1.007 | 1.005 | 1.005 | 1.002 | 1.000 | 1.005 |
| $\alpha_{min}/\alpha_{ideal}$ for Fe1 | 0.981 | 0.977 | 0.978 | 0.981 | 0.981 | 0.987 | 0.992 | 0.993 |

| $\alpha_{max}/\alpha_{ideal}$ for Fe2 | 1.028 | 1.029 | 1.032 | 1.036 | 1.037 | 1.037 | 1.040 | 1.043 |
|---|---|---|---|---|---|---|---|---|
| $\alpha_{min}/\alpha_{ideal}$ for Fe2 | 0.938 | 0.938 | 0.939 | 0.942 | 0.944 | 0.950 | 0.953 | 0.964 |

Table 6SM. Bond angles for Fe2-Fe7 tetrahedrons in 5 ×5 ×1 supercell.

| $z$ | 0 | 0.1 | 0.5 | 1.0 | 1.1 | 1.2 | 1.4 | 1.7 | 2 |
|---|---|---|---|---|---|---|---|---|---|
| α min at Se(S) for Fe2 | 58.786 | 57.55 | 57.36 | 57.61 | 57.72 | 57.71 | 57.77 | 58.13 | 57.55 |
| α max at Se(S) for Fe2 | 60.771 | 62.61 | 62.77 | 62.36 | 62.23 | 62.44 | 62.32 | 61.85 | 61.72 |
| α min at Se(S) for Fe3 | 56.178 | 56.18 | 56.09 | 56.385 | 56.22 | 56.42 | 56.84 | 57.21 | 58.08 |
| α max at Se(S) for Fe3 | 64.220 | 64.19 | 64.20 | 63.699 | 63.98 | 63.53 | 63.30 | 62.91 | 62.02 |
| α min at Se(S) for Fe4 | 55.881 | 56.31 | 56.09 | 56.355 | 56.60 | 56.53 | 56.58 | 56.91 | 56.20 |
| α max at Se(S) for Fe4 | 64.654 | 64.11 | 64.32 | 64.132 | 64.08 | 64.03 | 64.07 | 64.03 | 63.97 |
| α min at Se(S) for Fe5 | 56.103 | 55.36 | 55.31 | 55.276 | 55.36 | 55.63 | 55.91 | 55.89 | 56.90 |
| α max at Se(S) for Fe5 | 63.886 | 62.28 | 62.39 | 62.352 | 62.41 | 62.60 | 62.07 | 62.23 | 62.23 |
| α min at Se(S) for Fe6 | 55.875 | 56.91 | 57.04 | 57.132 | 57.37 | 57.43 | 57.44 | 57.92 | 58.46 |
| α max at Se(S) for Fe6 | 64.169 | 63.31 | 63.22 | 63.297 | 63.19 | 63.01 | 62.91 | 63.00 | 62.91 |
| α min at Se(S) for Fe7 | 56.322 | 57.52 | 57.34 | 57.37 | 57.60 | 58.86 | 57.91 | 58.13 | 57.64 |
| α max at Se(S) for Fe7 | 63.651 | 62.70 | 62.71 | 62.54 | 62.29 | 61.60 | 62.21 | 61.66 | 61.44 |

# C. Differential scanning calorimetry

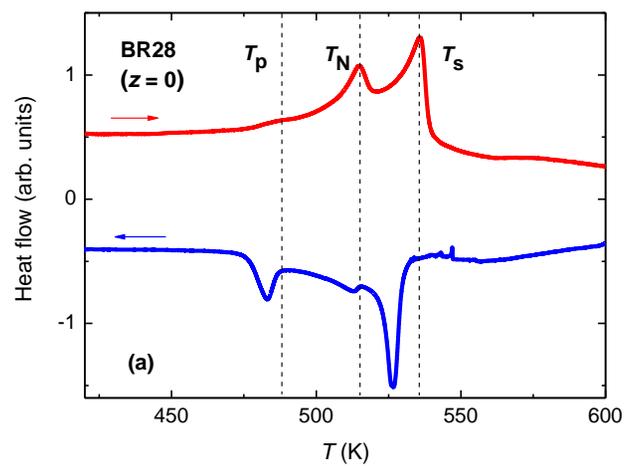

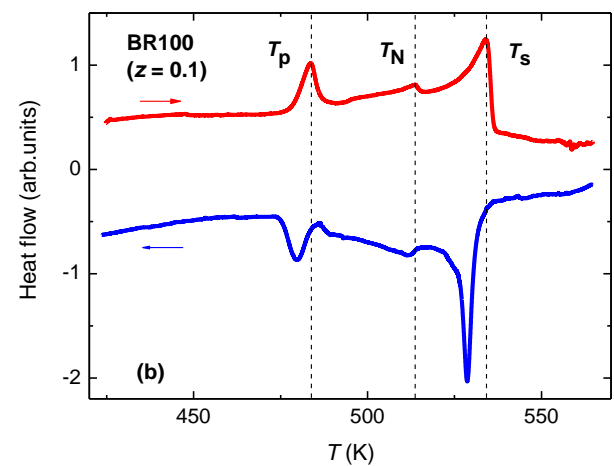

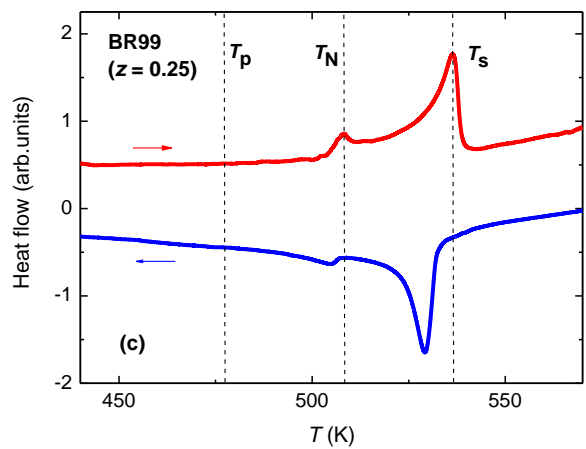

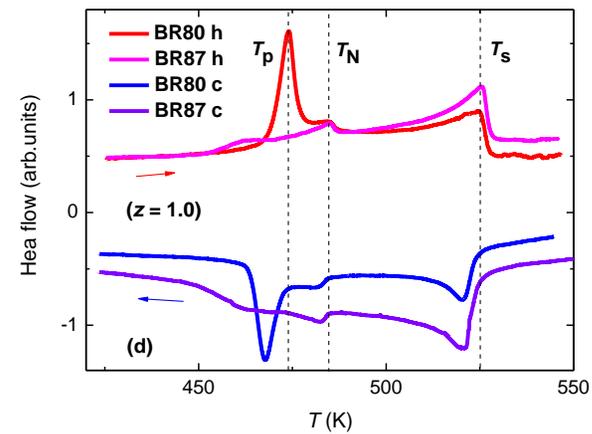

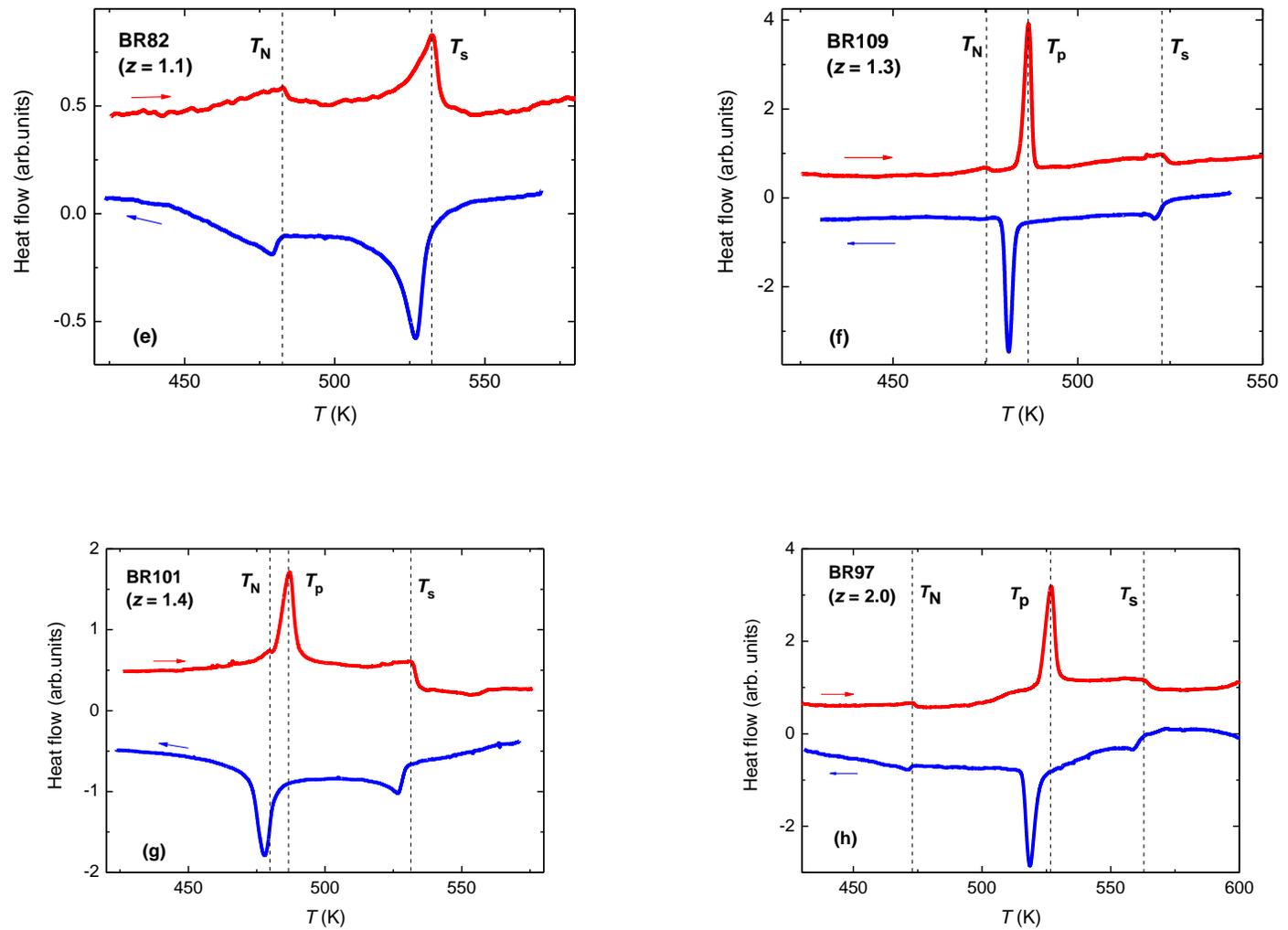

FIG. 4SM (a-h): Temperature dependence of DSC signals for $Rb_{1-x}Fe_{2-y}Se_{2-z}S_z$ crystals with different substitution. Red curves show data on heating, blue ones on cooling. Vertical dashed lines mark phase transformations on heating.

**D. Comparison of susceptibility and differential scanning calorimetry data**

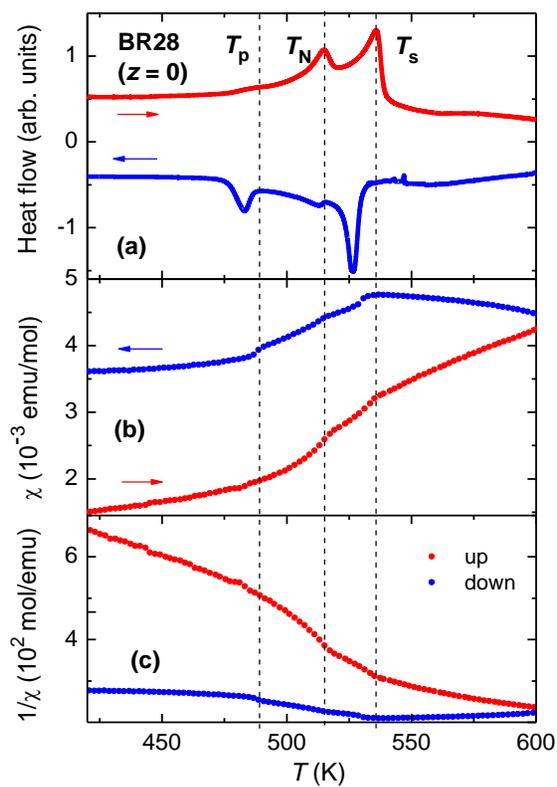

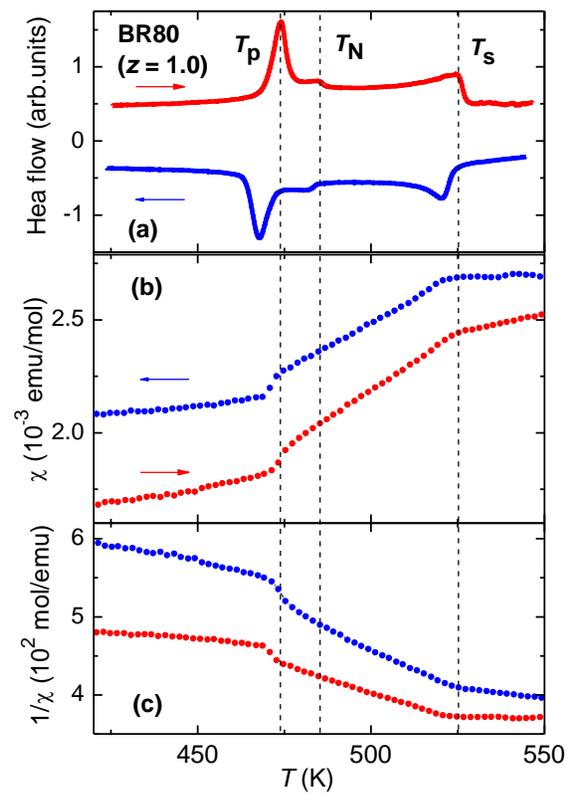

FIG. 5SM (a): (a) DSC signal, (b) susceptibility, and (c) inverse susceptibility *vs.* temperature for nonsubstituted sample ($z = 0$). Vertical dashed lines mark phase transformations on heating.

FIG. 5SM (b): (a) DSC signal, (b) susceptibility, and (c) inverse susceptibility *vs.* temperature for sample with substitution $z = 1.0$.

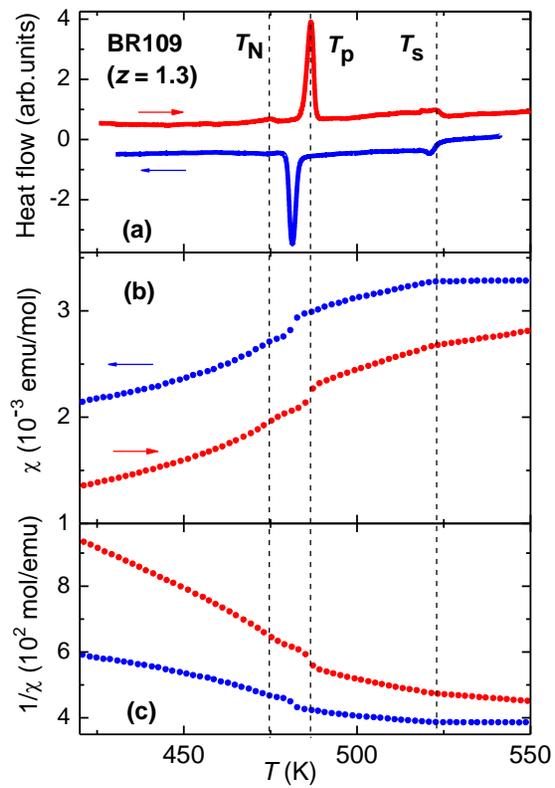

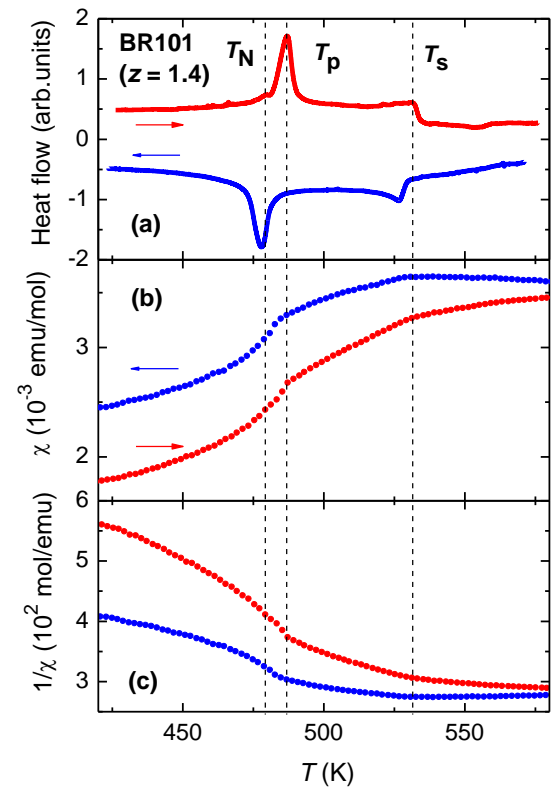

FIG. 5SM (c): (a) DSC signal, (b) susceptibility, and (c) inverse susceptibility *vs.* temperature for sample with substitution $z = 1.3$.

FIG. 5SM (d): (a) DSC signal, (b) susceptibility, and (c) inverse susceptibility *vs.* temperature for sample with substitution $z = 1.4$.

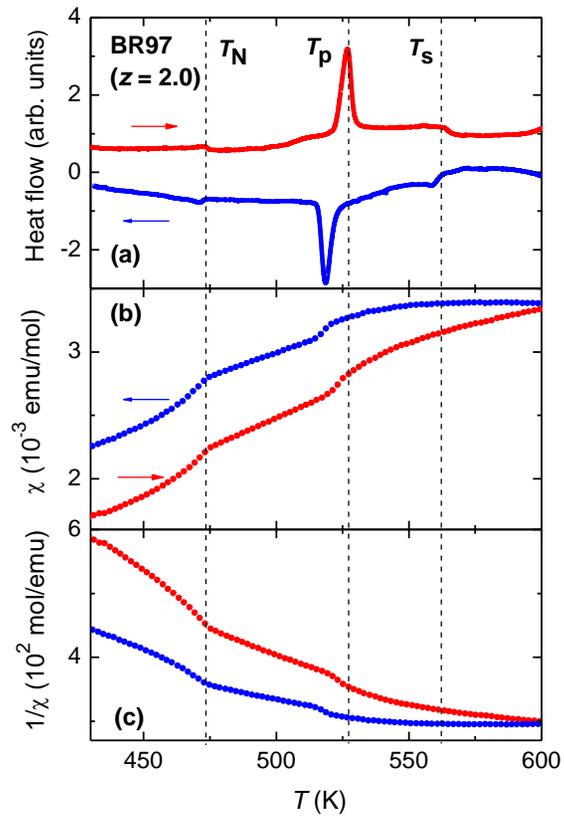

FIG. 5SM (e): (a) DSC signal, (b) susceptibility, and (c) inverse susceptibility *vs.* temperature for sample with substitution $z = 2.0$.

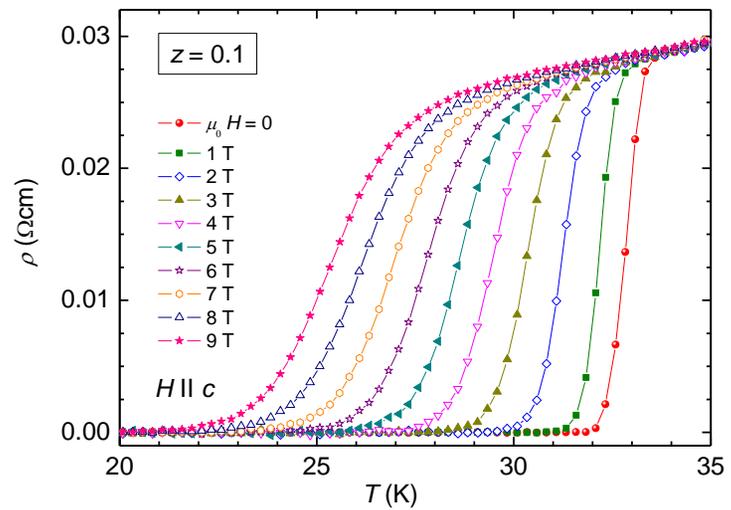
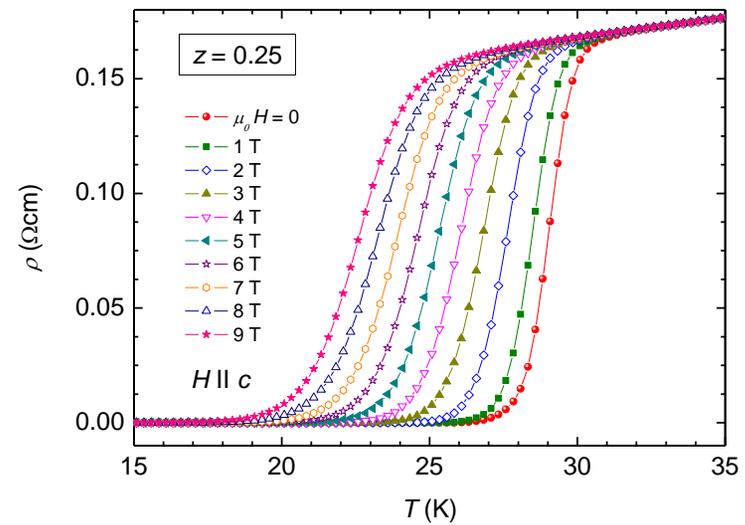

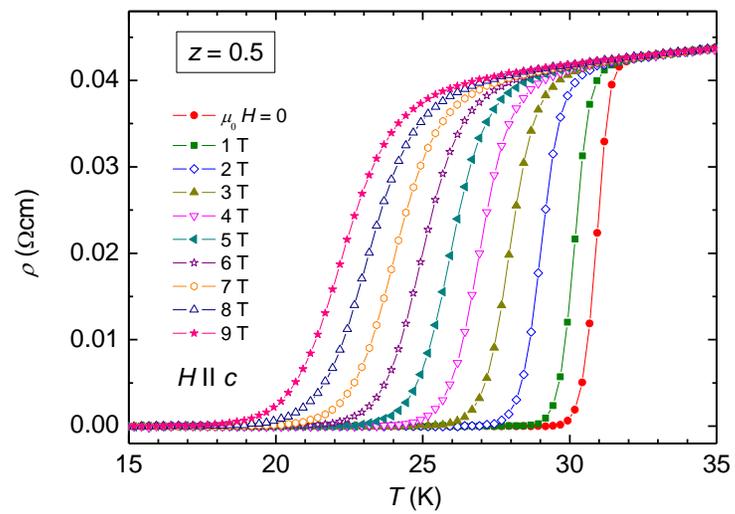
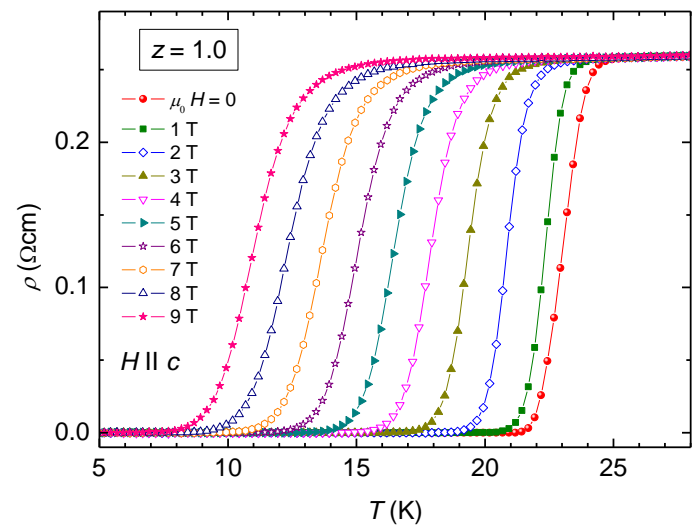

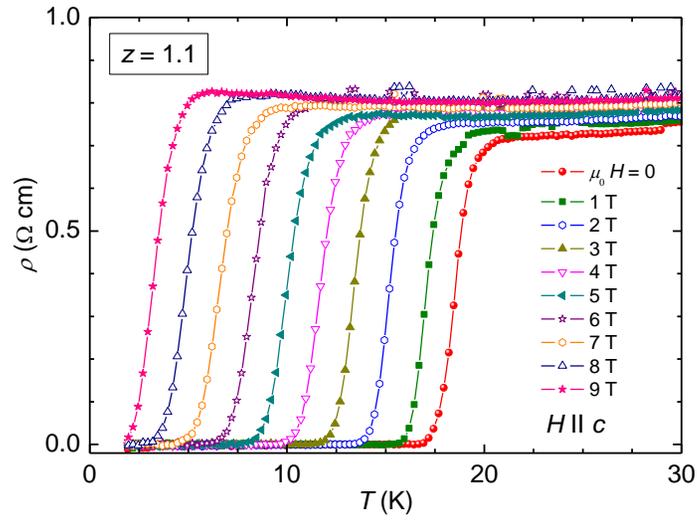 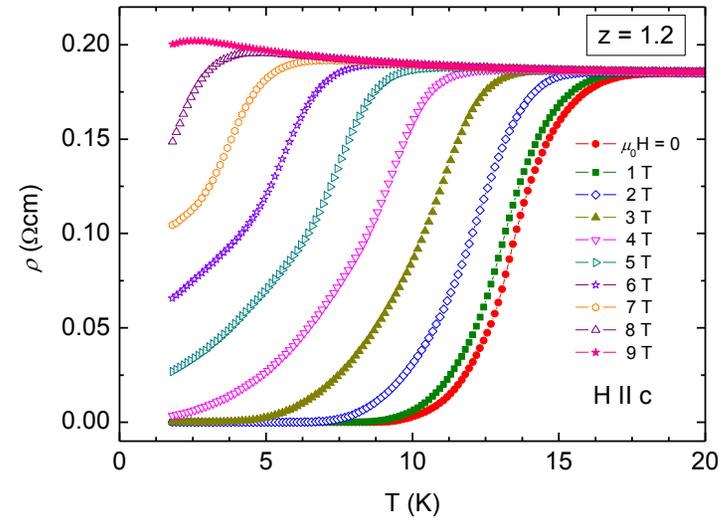

FIG. 6SM. Temperature dependent resistivity in different applied magnetic fields in the vicinity of superconducting transition for $Rb_{1-x}Fe_{2-y}Se_{2-z}S_z$ with $z$ = 0.1, 0.25, 0.5, 1.0, 1.1, and 1.2.